\documentclass[%
 reprint,
 amsmath,amssymb,
 aps,
]{revtex4-2}

\usepackage{graphicx}% Include figure files
\usepackage{dcolumn}% Align table columns on decimal point
\usepackage{bm}% bold math
\usepackage{enumerate}   
\usepackage{lineno}
\begin{document}

\preprint{APS/123-QED}

\title{Bayesian estimation of the specific shear and bulk viscosity of the quark-gluon plasma with additional flow harmonic observables}% Force line breaks with \\

%\email{anna.onnerstad@cern.ch}
\author{~J.E.~Parkkila$^{1,2}$}%
\email{jasper.parkkila@cern.ch}
\author{A.~Onnerstad$^{1,2}$}
\author{~D.J.~Kim$^{1,2}$}%
%\email{ dong.jo.kim@jyu.fi}
\affiliation{$^{1}$University of Jyv\"askyl\"a, Department of Physics, P.O. Box 35, FI-40014 University of Jyv\"askyl\"a, Finland}
\affiliation{$^{2}$Helsinki Institute of Physics, P.O.Box 64, FI-00014 University of
Helsinki, Finland}


%\collaboration{CLEO Collaboration}%

\date{\today}% It is always \today, today,

\begin{abstract}
The transport properties of the strongly-coupled quark-gluon plasma created in ultra-relativistic heavy-ion collisions are extracted by Bayesian parameter estimate methods with the latest collision beam energy data from the CERN Large Hadron Collider. This Bayesian analysis includes sophisticated flow harmonic observables for the first time. We found that the temperature dependence of specific shear viscosity appears weaker than in the previous studies. The results prefer a lower value of specific bulk viscosity and a higher switching temperature to reproduce additional observables. However, the improved statistical uncertainties both on the experimental data and hydrodynamic calculations with additional observables do not help to reduce the final credibility ranges much, indicating a need for improving the dynamical collision model before the hydrodynamic takes place. In addition, the sensitivities of experimental observables to the parameters in hydrodynamic model calculations are quantified. It is found that the analysis benefits most from the symmetric cumulants and non-linear flow modes, which mostly reflect non-linear hydrodynamic responses, in constraining the temperature dependence of the specific shear and bulk viscosity in addition to the previously used flow coefficients.

\end{abstract}

\maketitle

%\tableofcontents

\renewcommand{\vec}[1]{\mathbf{#1}}

\newcommand{\trento}{T\raisebox{-.5ex}{R}ENTo}

\section{\label{sec:intro}Introduction}
The primary goal of heavy-ion physics is to investigate and understand the strongly coupled color-deconfined matter, quark-gluon plasma (QGP), which is produced in ultrarelativistic collisions between heavy ions. The QGP is believed to be the predominant form of matter during the first phases of the early universe. This matter behaves like a near-perfect fluid with the smallest specific shear viscosity, the ratio of the shear viscosity to the entropy density ($\eta/s$), of any known substance in nature~\cite{Kovtun:2004de}.

The most important remaining open questions in the field are the location of the critical point ($T_c$) in the QCD phase diagram and temperature dependence of specific shear ($\eta$) and bulk ($\zeta$) viscosities of the QGP. The flow analysis in the Large Hadron Collider (LHC) has been very successful and provides valuable information to the field~\cite{ALICE:2011ab,ALICE:2016kpq,Acharya:2017gsw,Acharya:2020taj,ALICE:2021adw}. For example, the main constraints for the QGP properties using the Bayesian analysis~\cite{Bernhard2019} in the theory came from the ALICE measurements~\cite{Aamodt:2010cz, Abelev:2013vea, ALICE:2011ab} with both low and high beam energy data. 
Even though the Bayesian analyses~\cite{Bernhard2019,Auvinen:2020mpc,Nijs:2020ors,Nijs:2020roc,JETSCAPE:2020mzn} were successful, the current uncertainties from these works are large because of statistical limitations of the data, limited observables used for the analysis, and computational constraints. In addition to the aforementioned limitations, pinning down the absolute value of $\eta/s$ at $T_{c}$ has a few challenges. First, a principle calculation to describe the initial conditions (IC) is still under development. Second, extracting the temperature dependence of $\eta/s(T)$ has been complicated with the existence of the bulk viscosity~\cite{Ryu:2017qzn, Denicol:2009am}. However, large flow found in small systems like proton-proton (pp) collisions was striking and opened up the importance of gluon fluctuations within protons and certainly, the experimental data would help to improve the understanding of IC both for small and large systems~\cite{Acharya:2019vdf}. There are newer observables that give much stronger constraints to the theory~\cite{ALICE:2016kpq,Acharya:2017gsw}, showing good sensitivities especially to $\eta/s(T)$ and $\zeta/s(T)$. The correlation strength measured in~\cite{ALICE:2016kpq,Acharya:2017gsw} was experimentally decomposed into two components of linear and non-linear flow modes in~\cite{Acharya:2017zfg,Acharya:2020taj} for the first time in the field, which gives a better understanding of our harmonic analysis and its origin with both LHC Run 1 (2009-2013) and Run 2 (2015-2018) data. 

In this work, we extend the Bayesian parameter estimation methods employed in~\cite{Bernhard2019} with larger statistic LHC Run 2 results~\cite{Adam:2016ddh, Acharya:2019yoi} as well as a few additional observables~\cite{Acharya:2020taj, ALICE:2021adw} for the first time which require substantial computational power. This work also allows us to quantify the sensitivity of each observable to the hydrodynamic model parameters in a controlled way. In Sec.~\ref{sec:bayesian}, we present a brief overview of Bayesian analysis methods and model setups. The experimental observables are described in Sec.~\ref{sec:obs}. Model parameters and calibrations are explained in Sec.~\ref{sec:paramest_overview}. The results are presented in Sec.~\ref{sec:results}, after which Sec.~\ref{sec:summary} summarizes our results and findings.

\section{\label{sec:bayesian}Bayesian analysis}
There have been a number of studies that utilized Bayesian methods for heavy ion collisions~\cite{Petersen:2010zt,Novak:2013bqa,Pratt:2015zsa,Sangaline:2015isa,Heinz:2015arc}. We employ the recent state-of-the-art development in~\cite{Bernhard2019} for our present study. We define a vector of model parameters $\vec{x}$, and a set of experimental data $\vec{y}$ that will be compared with model calculations. Bayes's theorem gives the posterior distributions for the model parameters as
\begin{equation}
    P(\vec{x}|\vec{y})\propto P(\vec{y}|\vec{x}) P(\vec{x}).
\end{equation}
Here $P(\vec{y}|\vec{x})$ is the likelihood, which quantifies the model agreement with the data. The prior $P(\vec{x})$ encapsulates initial knowledge on the parameters.

The model parameters are then extracted from the posterior distributions. 
We follow the same procedures as~\cite{Bernhard2019}, where the model is first evaluated at a small $O(10^2)$ number of `design' parameter points. The resulting discrete set of model predictions is then made continuous by the use of a Gaussian process (GP) emulator, which thereby can be used to systematically probe the parameter space with Markov chain Monte-Carlo (MCMC) methods.

\subsection{Hydrodynamic model and parameters}
The model used in this analysis consists of the T$_{\rm{R}}$ENTo model~\cite{Moreland:2014oya} for the initial condition, which is connected with free streaming to a 2+1 dimensional causal hydrodynamic model VISH(2+1)~\cite{Shen:2014vra,Song:2007ux}. The evolution continues after particlization via the UrQMD model~\cite{Bass:1998ca,Bleicher:1999xi}. This hybrid model denoted \trento{}+VISH(2+1)+UrQMD, has successfully described the previous ALICE measurements~\cite{Bernhard2019}.

A hydrodynamic modeling relies on the energy and momentum conservation laws of the fluid dynamics. The conservation is expressed in terms of
\begin{equation}
	\label{eq:hydro1}
	\partial_\mu T^{\mu\nu}(x)=0,
\end{equation}
where $T^{\mu\nu}(x)$ is the energy-momentum tensor. In the case of viscous hydrodynamics, the energy-momentum tensor becomes
\begin{equation}
	\label{eq:hydro2}
	T^{\mu\nu}=\epsilon u^\mu u^\nu-(P+\Pi)\Delta^{\mu\nu}+\pi^{\mu\nu},
\end{equation}
where $\epsilon$ is the energy density, $P$ is the local pressure given by the equation of state, and $\Delta^{\mu\nu}=g^{\mu\nu}-u^\mu u^\nu$ is a projector onto the transverse four-velocity. The shear- and bulk viscosities are encoded into $\pi^{\mu\nu}$ and $\Pi$, respectively.

Free parameters of this model include the initial conditions, $\eta/s(T)$ and $\zeta/s(T)$, characterized by a total of 14 model parameters, which together control the prominent features of the model. The parameter set, described in detail in later sections, will enable simultaneous characterization of the initial state and medium response, including any correlations.

Each event consists of a single initial condition given energy density profile and a hydrodynamic simulation followed by multiple samples of the freeze-out hypersurface.
The parameter estimation is conducted using 500 parameter design points, sampled evenly from the parameter space using the Latin hypercube scheme~\cite{TangHypercube,MORRIS1995381}. At each design point, the model is used to generate around 3$\times 10^{5}$ events with the corresponding parametrization, with each event surface sampled ten times to produce a total of $3\times 10^6$ events for 0--60\% centrality ranges. A large number of events is generated to ensure a better accuracy for high harmonic observables. A GP emulator is then trained to produce predictions for the observables in between the design points, after which the predictions are validated against a validation set. See~\cite{Bernhard2019} for details of the emulator. Using the emulator to produce predictions in continuous parameter space, the final posterior distribution is created using MCMC sampling.

\subsection{Calibrating the model parameters}
\newcommand{\x}{\mathbf x}
\newcommand{\y}{\mathbf y}
\newcommand{\z}{\mathbf z}
\newcommand{\st}{_\star}
\newcommand{\ex}{_\text{exp}}

The parameter estimation attempts to calibrate the model parameters for the model to optimally reproduce experimental observables. With Bayesian methods, the optimal parameters are characterized by probability distributions for their true values.
As given by Bayes's theorem, the probability for the true parameters $\x\st$ is
\begin{equation}
  P(\x\st|X,Y,\y\ex) \propto P(X,Y,\y\ex|\x\st) P(\x\st).
  \label{eq:bayes}
\end{equation}
The left-hand side is the \emph{posterior}: the probability of $\x\st$ given the design $X$, computed observables $Y$, and the experimental data $\y\ex$.
On the right-hand side, $P(\x\st)$ is the \emph{prior} probability, encapsulating the initial knowledge of $\x\st$, and $P(X,Y,\y\ex|\x\st)$ is the likelihood: the probability of observing $(X, Y, \y\ex)$ given a proposal $\x\st$.

The likelihood may be computed using the principal component GP emulators as
\begin{align}
  P &= P(X,Y,\y\ex|\x\st) \nonumber \\
    &= P(X,Z,\z\ex|\x\st) \nonumber \\
    &\propto\exp\biggl\{
      -\frac{1}{2} (\z\st - \z\ex)^\top \Sigma_z^{-1} (\z\st - \z\ex)
    \biggr\},
  \label{eq:likelihood}
\end{align}
where $\z\st = \z\st(\x\st)$ are the principal components predicted by the emulators, $\z\ex$ is the principal component transform of the experimental data $\y\ex$, and $\Sigma_z$ is the covariance (uncertainty) matrix.
The covariance matrix encodes all the experimental and model uncertainties~\cite{Bernhard:2018hnz}. In the principal component space, the covariance matrix can be expressed as
\begin{equation}
    \Sigma_z=\Sigma_z^\mathrm{exp}+\Sigma_z^\mathrm{GP}+(\sigma_m^\mathrm{sys})^2 I,
\end{equation}
where $\Sigma_z^\mathrm{exp}$ is the matrix for experimental errors and $\Sigma_z^\mathrm{GP}=\mathrm{diag}(\sigma_{z,1}(\z\st)^2,\sigma_{z,2}(\z\st)^2,\dots,\sigma_{z,k}(\z\st)^2)$ is the diagonal GP emulator covariance matrix, representing the model statistical and GP predictive uncertainty. Additionally, $\sigma_m^\mathrm{sys}$ is a free parameter ranging from zero to one, with the purpose of including all remaining uncertainties arising from the model imperfections.
All model parameters are given a uniform prior. Together with the likelihood~\eqref{eq:likelihood} and Bayes' theorem~\eqref{eq:bayes}, the posterior probability can evaluated at an arbitrary point in the parameter space. To construct the posterior distribution, an MCMC method can be used, which generates random walks through parameter space by accepting or rejecting proposal points based on the posterior probability.

\section{\label{sec:obs}Experimental Observables}
In the previous studies, the centrality dependence of identified particle yields $\mathrm{d}N/\mathrm{d}y$ and mean transverse momenta $\langle p_\mathrm{T}\rangle$ for charged pions, kaons, and protons as well as two-particle anisotropic flow coefficients $v_n$ for $n = 2$, 3, 4 were used. The observables are measured by the ALICE Collaboration in Pb--Pb collisions at $\sqrt{s_\mathrm{NN}} = 2.76$~TeV~\cite{Aamodt:2010cz, Abelev:2013vea, ALICE:2011ab}.

\begin{table*}[tbh!]
  \caption{
    \label{tab:observables}
    Experimental data included in Bayesian analysis.
  }
  %\begin{ruledtabular}
  \begin{tabular}{|lcccc|}
    \hline
    Observable & Particle species & Kinematic cuts & Centrality classes & Ref. \\
    %\paddedhline
    \hline
    Yields $\mathrm{d}N/\mathrm{d}y$                       & $h^\pm$, $p\bar p$ &
    $|y| < 0.5$ & 0--5, 5--10, 10--20, \ldots, 50--60 & \cite{Adam:2016ddh} \\
    %\noalign{\smallskip}
    Mean transverse momentum $\langle p_\mathrm{T}\rangle$ & $\pi^\pm$, $K^\pm$, $p\bar p$ &
    $|y| < 0.5$ & 0--5, 5--10, 10--20, \ldots, 50--60 & \cite{Acharya:2019yoi} \\
    %\noalign{\smallskip}
    Two-particle flow cumulants & $h^\pm$ &
    $|\eta| < 0.8$ & 0--5, 5--10, 10--20, \ldots, 50--60 &
    \cite{ Acharya:2020taj} \\
    $n = 2-8$ & & $0.2 <  p_\mathrm{T}< 5.0$ GeV &  & \\
    %\noalign{\smallskip}
    Non-linear flow mode & $h^\pm$ &
    $|\eta| < 0.8$ & 0--5, 5--10, 10--20, \ldots, 50--60 &
    \cite{Acharya:2020taj} \\
    $n = 2-8$ & & $0.2 < p_\mathrm{T} < 5.0$ GeV &  & \\
    Symmetric cumulants & $h^\pm$ &
    $|\eta| < 0.8$ & 0--5, 5--10, 10--20, \ldots, 50--60 &
    \cite{ALICE:2021adw} \\
    $n = 2-8$ & & $0.2 <  p_\mathrm{T} < 5.0$ GeV &  & \\
    \hline
  \end{tabular}
  %\end{ruledtabular}
\end{table*}

\begin{figure}[tbh!]
	\centering
	\includegraphics[width=0.9\linewidth]{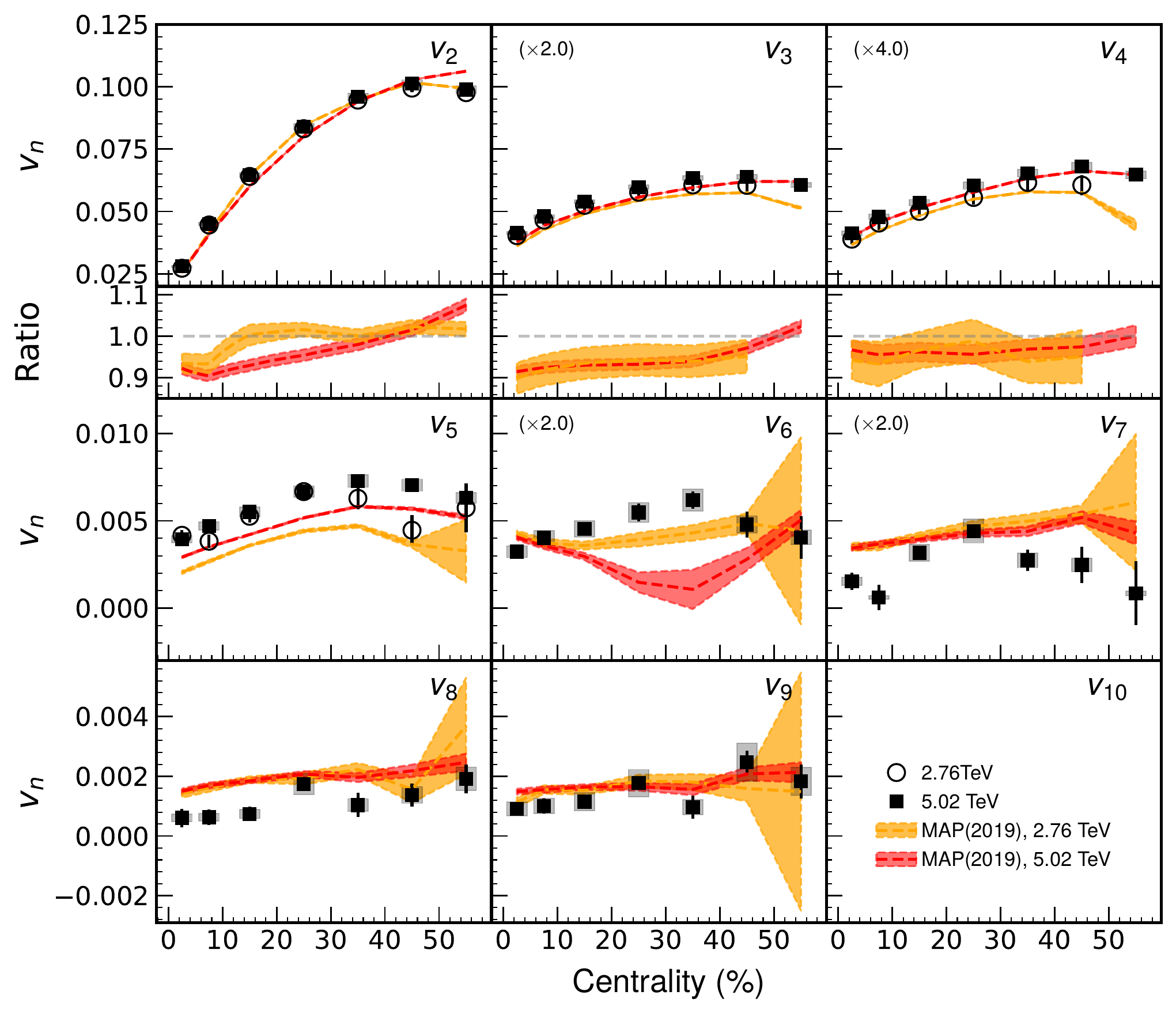}
	\caption{Model calculations of flow coefficients compared to experimental data at center-of-mass energies of 2.76 and 5.02~TeV. The systematic error for the higher energy data points is shown as a grey band around them. This band is not shown for the lower energy data points since they have combined errors.}
	\label{fig:vn_nat}
\end{figure}

\begin{figure}[tbh!]
	\centering
	\includegraphics[width=0.9\linewidth]{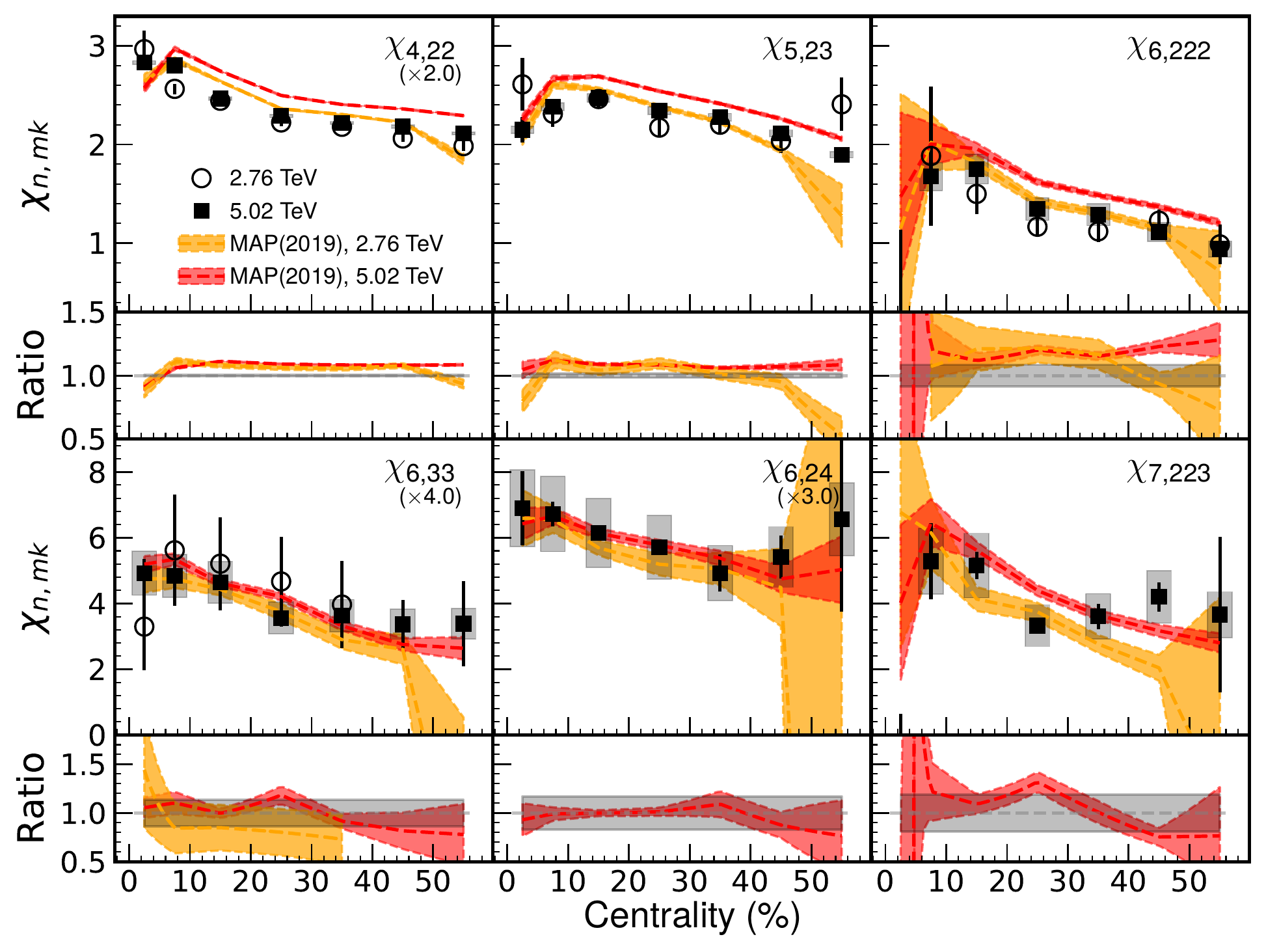}
	\caption{Model calculations of non-linear flow mode coefficients compared to experimental data. Most calculations reproduce $\chi_{4,22}$ within the uncertainties of the measurement and calculations. The systematic error for the higher energy data points is shown as a grey band around them. This band is not shown for the lower energy data points since they have combined errors.}
	\label{fig:chi_nat}
\end{figure}

\begin{figure}[tbh!]
	\centering
	\includegraphics[width=0.9\linewidth]{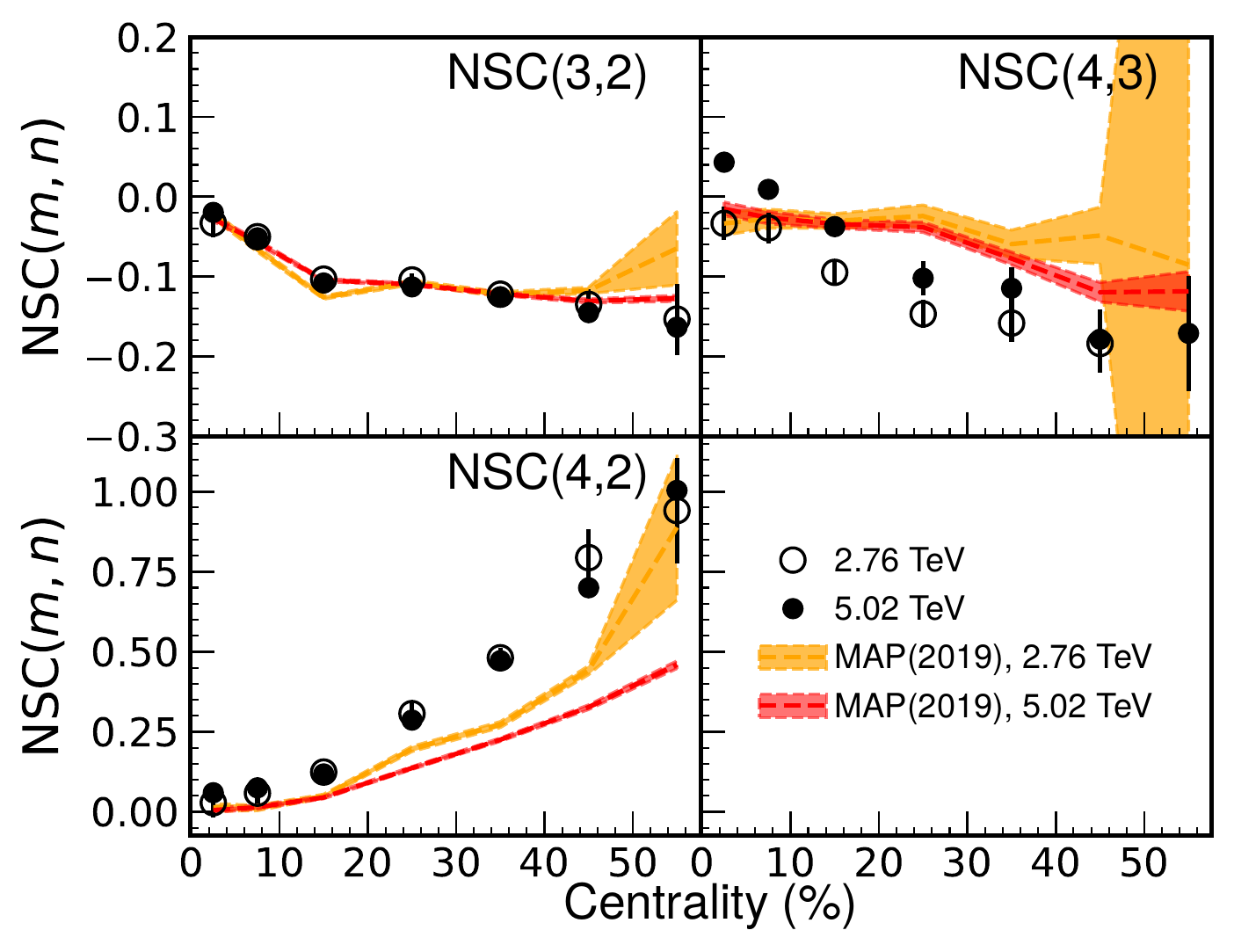}
	\caption{Model calculations of the normalized symmetric cumulants [$\mathrm{NSC}(m,n)$] compared to experimental data at center-of-mass energies of 2.76 and 5.02~TeV. The systematic error for the higher energy data points is shown as a grey band around them. This band is not shown for the lower energy data points since they have combined errors.}
	\label{fig:nsc_nat}
\end{figure}

\begin{figure}[tbh!]
	\centering
	\includegraphics[width=1.0\linewidth]{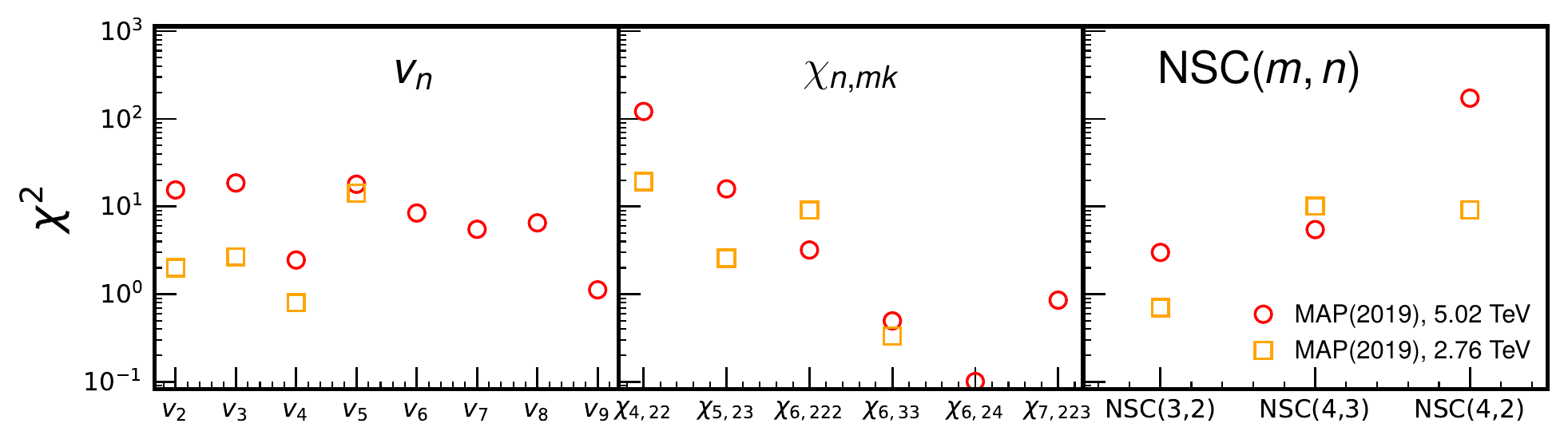}
	\caption{The $\chi^2$ values calculated between the data and model calculations for both beam energies are shown for $v_n$, $\chi_{n,mk}$, and $\mathrm{NSC}(m,n)$.}
	\label{fig:chi_squared}
\end{figure}

In this work, we mainly focus on the larger statistic higher beam energy collisions at $\sqrt{s_\mathrm{NN}} = 5.02$~TeV, which give better precision. In addition to the above mentioned observables, we include higher harmonic flow coefficients $v_n$~\cite{Acharya:2020taj} (up to $n=9$), the normalized symmetric cumulants, $\mathrm{NSC}(m,n)$~\cite{ALICE:2021adw}, and the non linear flow mode coefficients $\chi_{n,mk}$~\cite{Acharya:2020taj}. 

The anisotropic pressure-driven expansion of the QGP, commonly referred to as anisotropic flow, can be characterized by a Fourier decomposition of the azimuthal particle distributions as 
\begin{equation}
   \frac{\mathrm{d}N}{\mathrm{d}\phi} \propto 1 + 2 \sum_{n=1}^{\infty}v_n \cos\left(n(\phi-\psi_{n})\right), 
   \label{eq:fourier}
\end{equation} 
where $v_n$ quantifies the magnitude of the $n^\mathrm{th}$ harmonic flow, and $\psi_{n}$ its direction.
$\mathrm{NSC}(m,n)$  quantifies the correlations between event-by-event fluctuations of flow harmonics of different orders, $\mathrm{NSC}(m,n) = (\left<v_{m}^2v_{n}^2\right>-\left<v_{m}^2\right>\left<v_{n}^2\right>)/\left<v_{m}^2\right>\left<v_{n}^2\right>$~\cite{ALICE:2016kpq,ALICE:2021adw} and $\chi_{n,mk}$ measures the contribution of lower order harmonic flows to higher order harmonics (i.e., $\chi_{4,22}$ is the non linear contribution of $v_4$ originating from $v_2$; see the details in~\cite{Acharya:2020taj}).
These additional observables give better sensitivity to the medium properties and initial conditions as demonstrated in Refs.~\cite{ALICE:2016kpq,Acharya:2017gsw,Acharya:2020taj,ALICE:2021adw}.

As shown in Fig.~\ref{fig:vn_nat} and \ref{fig:chi_nat}, a model calculation with the best-fit parametrization given by maximum a posteriori (MAP) from the previous Bayesian analysis~\cite{Bernhard2019} shows deviations of the measurements for the flow coefficients from $n=5$ and the non-linear flow mode coefficients from $\chi_{n,mk}$ ($n=4$).
The black filled and open circles represent the higher ($\sqrt{s_\mathrm{NN}} = 5.02\,\mathrm{TeV}$) and lower ($\sqrt{s_\mathrm{NN}} = 2.76\,\mathrm{TeV}$) energy data points, respectively, whereas the red and orange bands represent the higher and lower energy model calculations. The $v_2 - v_4$ values calculated from data were used in model calibration, and as seen in Fig.~\ref{fig:vn_nat}, the calculations agrees well for $v_4$. 
However, a discrepancy is seen for $v_3$ with an underestimation of the calculations for the centrality up to $\approx 45\%$ for both energies, and an even larger discrepancy in $v_2$ for the higher energy calculation while the lower energy calculation agrees well except for the low centrality of $0-10\%$.
For higher harmonics ($\geq v_5$) the deviation is still visible.

The model calculations for the non linear flow mode coefficients in Fig.~\ref{fig:chi_nat} agree within $\pm 15\%$ for $\chi_{4,22}$ and $+15\%$ for the higher energy model calculation of $\chi_{5,23}$, while the lower energy model calculation goes to -30\% in central collisions and even larger than 50\% at high centrality ranges. 
The discrepancies between data and model calculations are significantly larger from $\chi_{6,222}$; however, for $\chi_{6,24}$, most of the higher energy data points agree with the calculations within systematic uncertainties. 

Model calculations reproduce the value for $\mathrm{NSC}(3,2)$ up to the $40-50\%$ centrality class, which is shown in Fig.~\ref{fig:nsc_nat}.
Both model predictions underestimate the values of $\mathrm{NSC}(4,2)$ for all centrality classes presented.
The model calculations overestimate $\mathrm{NSC}(4,3)$ for the lower energy data and give similar results for the higher energy. However, the results show clear differences between the two beam energies. The differences get larger toward the central collisions. While it is negative for the lower energy data and the model calculations, the measurement at 5.02~TeV shows the change of the signs in central collisions. Also, the magnitudes are smaller in lower energy collisions, which is attributed to the increasing contribution from the nonlinear hydrodynamic response in $v_4$~\cite{ALICE:2021adw}.

In order to quantify the agreement of the models with the data, the $\chi^2$ test was performed in the same way as in Eq.~(5) in Ref~\cite{Acharya:2017gsw} for the centrality range 5--50\%. The results are shown in Fig.~\ref{fig:chi_squared} for the flow coefficients, non-linear flow mode coefficients, and the normalized symmetric cumulants.
A significant difference is observed between the $\chi^2$ values for $v_n$ of higher and lower energies at $n\leq 4$.
%When comparing the $\chi^2$ values for $v_n$ of the higher energy to the lower energy we can see that there is a significant difference between them for $v_2$, $v_3$, and $v_4$, with the higher energy having a larger $\chi^2$ value. 
The $\chi^2$ values for $v_5$ are larger for both beam energies with similar magnitudes.
The higher energy $\chi^2$ value for $\chi_{4,22}$ is significantly larger than the one from the lower energy as shown in Fig.~\ref{fig:chi_squared}. 
The disagreement is still significant for $\chi_{5,23}$ and $\chi_{6,222}$.
For $\mathrm{NSC}(m,n)$, the $\chi^2$ values are larger for higher harmonics at both beam energies. The $\chi^2$ is especially large for the higher beam energy $\mathrm{NSC}(4,3)$.

In our calculations of the observables, we used the same methods also used in experimental analysis in Refs.~\cite{ALICE:2016kpq,Acharya:2017gsw,Acharya:2020taj,ALICE:2021adw}. Our centrality classes in this study are chosen to match those used for the experimental data. We define the multiplicity range for each centrality class by simulating events using the MAP-parametrization from~\cite{Bernhard2019}, and sorting the resulting minimum-bias events by charged-particle multiplicity $\mathrm{d}N/\mathrm{d}\eta$ at midrapidity ($|\eta| < 0.5$). The identified $\mathrm{d}N/\mathrm{d}\eta$ and $\langle p_\mathrm{T}\rangle$ were evaluated by counting and averaging the particle species at midrapidity ($|y| < 0.5$). The experimental data are readily corrected and extrapolated to zero $p_\mathrm{T}$~\cite{Abelev:2013vea}, and therefore no additional processing is required while preparing the comparison.
%The same centrality classes were selected as the corresponding experimental data by sorting each design point's minimum-bias events by charged-particle multiplicity $\mathrm{d}N/\mathrm{d}\eta$ at midrapidity ($|\eta| < 0.5$) and dividing the events into the desired percentile bins.
%The identified $\mathrm{d}N/\mathrm{d}\eta$ and $\langle p_T\rangle$ were computed by simple counting and averaging of the desired species at midrapidity ($|y| < 0.5$); no additional steps are necessary since the experimental data are corrected and extrapolated to zero $p_T$~\cite{Abelev:2013vea}.
For the identified $\mathrm{d}N/\mathrm{d}\eta$, only protons were used in model calibration, as the model did not reproduce the spectra of the other species with any of the parametrizations.
Finally, we calculated flow coefficients and other observables for charged particles within the kinematic range of the ALICE detector using the same methods as in the data analyses~\cite{Acharya:2020taj, ALICE:2021adw}.
A summary of all the observables that are included in the Bayesian analysis is given in 
Table~\ref{tab:observables}. The table presents the particle species, kinematic cuts, and centrality classes for each observable.

\section{\label{sec:paramest_overview}Parameter estimation using new LHC measurements}
\begin{table*}[tbh!]
  \caption{
    \label{tab:design}
    Input parameter ranges for the initial condition and hydrodynamic models.
  }
  \begin{tabular}{|llc|}
    \hline
    Parameter         & Description                        & Range    \\
    \hline
    Norm              & Overall normalization              & 16.542 -- 25  \\
    $p$               & Entropy deposition parameter       & 0.0042 -- 0.0098   \\
    $\sigma_k$ & Std. dev. of nucleon multiplicity fluctuations & 0.5508 -- 1.2852 \\
    $d_{\min}^3$ & Minimum volume per nucleon & $0.889^3$ -- $1.524^3$   \\
    $\tau_\mathrm{fs}$ & Free-streaming time & 0.03 -- 1.5  \\
    $T_c$ & Temperature of const. $\eta/s(T)$, $T < T_c$ & 0.135 -- 0.165 \\
    $\eta/s(T_c)$ & Minimum $\eta/s(T)$  & 0 -- 0.2  \\
    $(\eta/s)_\mathrm{slope}$ & Slope of $\eta/s(T)$ above $T_c$ & 0 -- 4 \\
    $(\eta/s)_\mathrm{crv}$ & Curvature of $\eta/s(T)$ above $T_c$ & $-1.3$ -- $1$  \\
    $(\zeta/s)_\mathrm{peak}$  & Temperature of $\zeta/s(T)$ maximum & 0.15 -- 0.2  \\
    $(\zeta/s)_{\max}$ & Maximum $\zeta/s(T)$ & 0 -- 0.1  \\
    $(\zeta/s)_\mathrm{width}$ & Width of $\zeta/s(T)$ peak & 0 -- 0.1  \\
    $T_\mathrm{switch}$ & Switching / particlization temperature & 0.135 -- 0.165  \\
    \hline
  \end{tabular}
\end{table*}

The model to be evaluated in this analysis consists of multiple stages, of which a brief overview will be given next. Altogether, the model setup includes the parametric \trento{} initial conditions, free-streaming pre-equilibrium dynamics, and the VISH(2+1) hydrodynamic model for medium evolution. Furthermore, the model performs the hadronization and includes UrQMD hadronic cascade. The model setup used is identical to the one developed and used in~\cite{Bernhard2019}, except for the number of hypersurface samples taken after evolution. In this work, exactly ten events are sampled from the hypersurface regardless of the cumulative number of particles. The centrality definition is shared for all parametrizations.
With close to fixed initial stage parameters, the possible effects of a shared centrality definition should be negligible.

Our main focus will be to investigate the effects of the higher harmonic observables on the temperature dependence of the transport coefficients. The parametrizations of the transport coefficients are~\cite{Bernhard2019}
\begin{equation}
	\label{eq:etaparam}
	(\eta/s)(T)=(\eta/s)(T_c)+(\eta/s)_\mathrm{slope}(T-T_c)\left(\frac{T}{T_c}\right)^{(\eta/s)_\mathrm{crv}}
\end{equation}
and
\begin{equation}
	\label{eq:zetaparam}
	(\zeta/s)(T)=\frac{(\zeta/s)_\mathrm{max}}{1 + \left(\frac{T-(\zeta/s)_{\mathrm{peak}}}{(\zeta/s)_\mathrm{width}}\right)^2}
\end{equation}
for the ratio of shear viscosity and bulk viscosity over entropy, respectively. Based on previous work, it is known that the lowest value of $\eta/s(T)$ is around the critical temperature $T_c$, close to the universal minimum $1/(4\pi)$. The temperature dependence of $\eta/s(T)$ is moderate, and increasing with higher values of temperature. Within close proximity of $150$ to $500\,\mathrm{MeV}$, the slope of $\eta/s(T)$ is approximately linear. The bulk viscosity over entropy ratio $\zeta/s(T)$ is expected to peak around $T_c$, and to decrease at higher values of temperature.

With this knowledge, we may construct our priori, and assume the initial parameter ranges. The chosen parameter ranges are loosely based on the optimal parameters found in~\cite{Bernhard2019}. It was found that in most cases, by taking the optimal parameters in~\cite{Bernhard2019} as the center points of the prior range and expanding the range slightly based on a reasonable $\sigma$ value, those parameters could be further optimized with the additional observables. In this study, we have kept the initial stage parameter ranges narrow around the MAP values found in~\cite{Bernhard2019} with the assumption that the additional observables affect mostly the transport coefficients. Very small variation was allowed to give the parameters space to adjust for minor differences.

The included and varied parameters, of which there are 14 in total, are summarized in Table~\ref{tab:design}. The parametric \trento{} initial conditions comprise an ansatz in terms of five parameters: a normalization factor $\mathrm{Norm}$, entropy deposition parameter $p$, standard deviation of the nuclear multiplicity fluctuations $\sigma_\mathrm{fluct}$, Gaussian-shaped nucleon width $w$, and minimum allowed distance between nucleons $d_\mathrm{min}$. The initial conditions are assumed to be already well constrained and presumably not affected by the addition of medium effect sensitive observables. The range for free-streaming time $\tau_\mathrm{fs}$ characterizing the allotted time for pre-equilibrium dynamics was kept relatively large.

\begin{figure}[tbh!]
	\centering
	\includegraphics[width=0.8\linewidth]{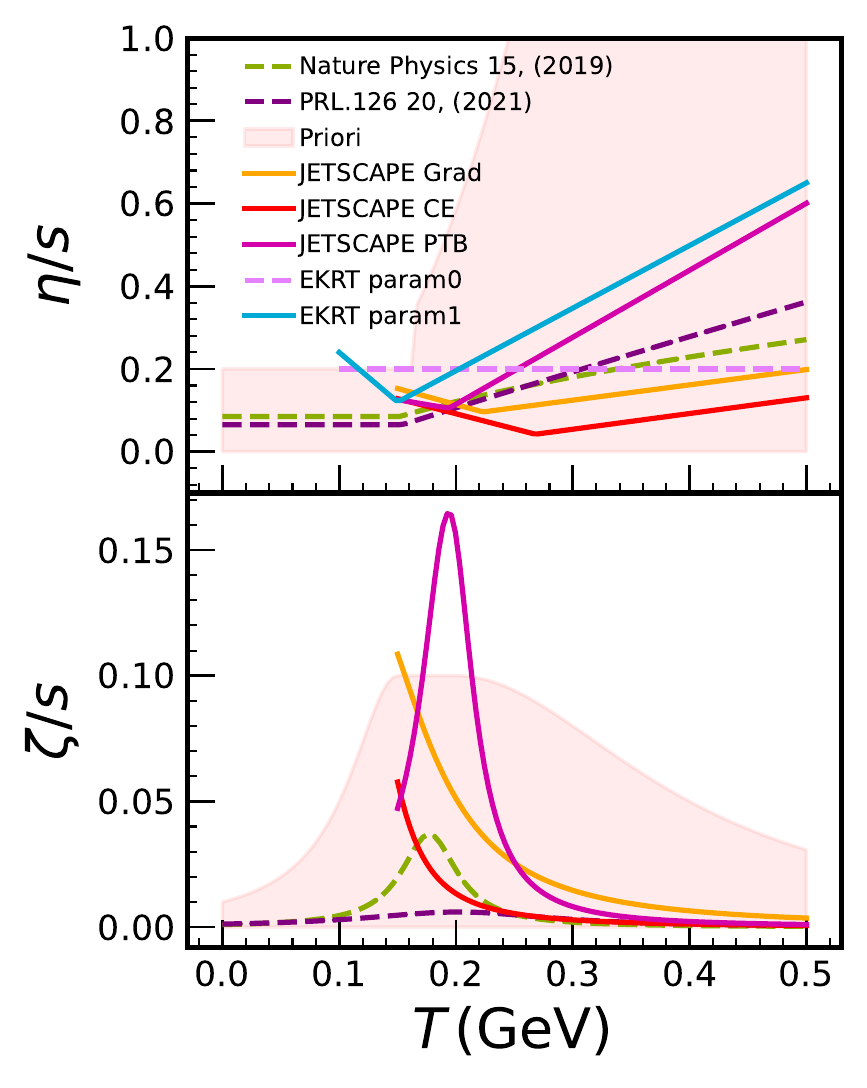}
	\caption{The specific shear ($\eta/s$) and bulk ($\zeta/s$) viscosity ratios as a function of temperature. The region plotted in red visualizes the prior range used in this study. Other curves represent some of the parametrizations found in previous studies: the best-fit $\eta/s(T)$ with EKRT initial conditions~\cite{Niemi:2015qia,Niemi:2015voa}, parametrizations from the JETSCAPE Collaboration with three different particlization distributions~\cite{JETSCAPE:2020mzn}, and a recent parametrization found in~\cite{Bernhard2019}.}
	\label{fig:prior}
\end{figure}

The rest of the parameters are the components of the transport coefficient parametrizations, and the switching temperature $T_\mathrm{switch}$ describing the temperature at which the hadronization begins to take place. The initial ranges given for these parameters are more generous, although large deviations in the final parameters compared to the previous study are not expected. The prior range for the transport coefficients is plotted in Fig.~\ref{fig:prior} among some parametrizations from other related studies~\cite{Niemi:2015qia,Niemi:2015voa,Bernhard2019,JETSCAPE:2020mzn}. The parametrizations are valid only up to the corresponding limits of the model; $100\,\mathrm{MeV}$ in the case of EKRT and $150\,\mathrm{MeV}$ for JETSCAPE. We note that the parametrizations EKRT+param0 and EKRT+param1 were not obtained through Bayesian analysis and we do not consider the slightly higher $\eta/s$ at around $T=100\,\mathrm{MeV}$ in our prior. Furthermore, we do not consider the large $\zeta/s$ reported with the PTB particlization by the JETSCAPE Collaboration~\cite{JETSCAPE:2020mzn}. Nevertheless, the $\zeta/s$ obtained using the Grad or CE particlization distributions are within our prior, considering the temperature limit $T_\mathrm{switch}>150\,\mathrm{MeV}$.

\begin{figure}[tbh!]
	\centering
	\includegraphics[width=1.0\linewidth]{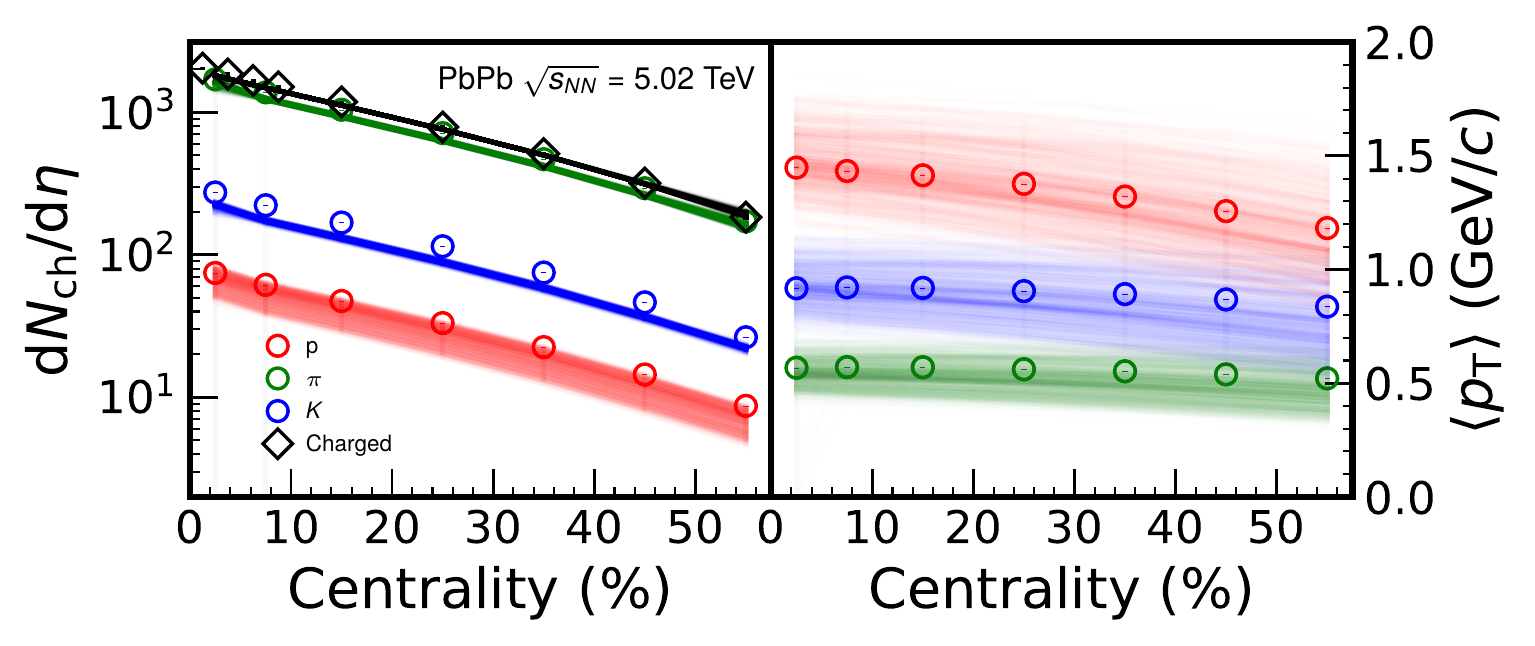}
	\caption{Charged and identified particle multiplicity and mean transverse momenta $\langle p_\mathrm{T}\rangle$ as given by the design parametrizations.}
	\label{fig:nch_meanpt_params}
\end{figure}

\begin{figure}[tbh!]
	\centering
		\includegraphics[width=0.8\linewidth]{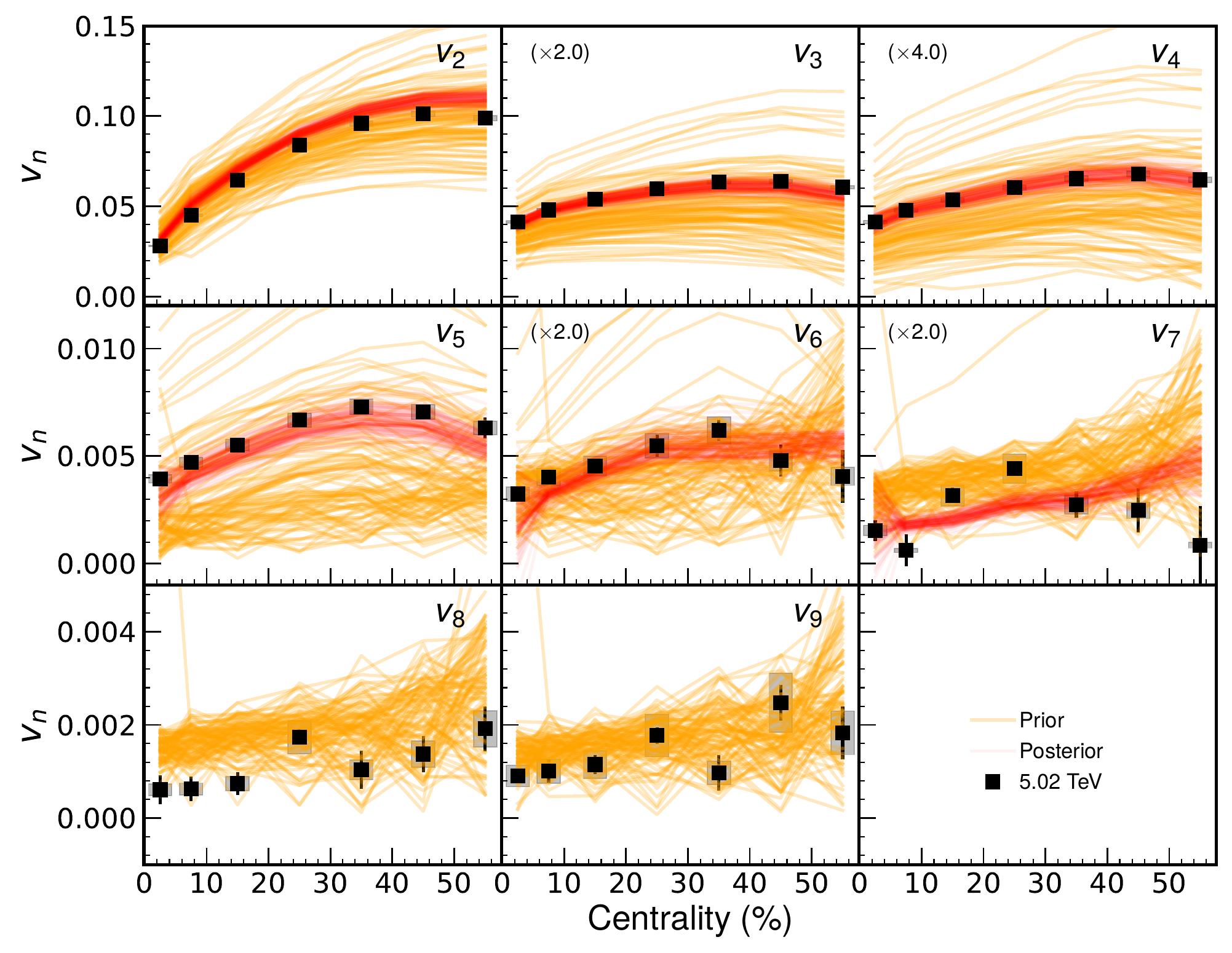}
	\caption{Flow coefficients $v_n$ as given by the design parametrizations are presented in yellow curves. All harmonics are simultaneously covered by the design parametrizations. The red curves represent a number of curves sampled from the posterior distribution, and as given by the emulator.}
	\label{fig:vn_params}
\end{figure}

\begin{figure}[tbh!]
	\centering
	\includegraphics[width=0.8\linewidth]{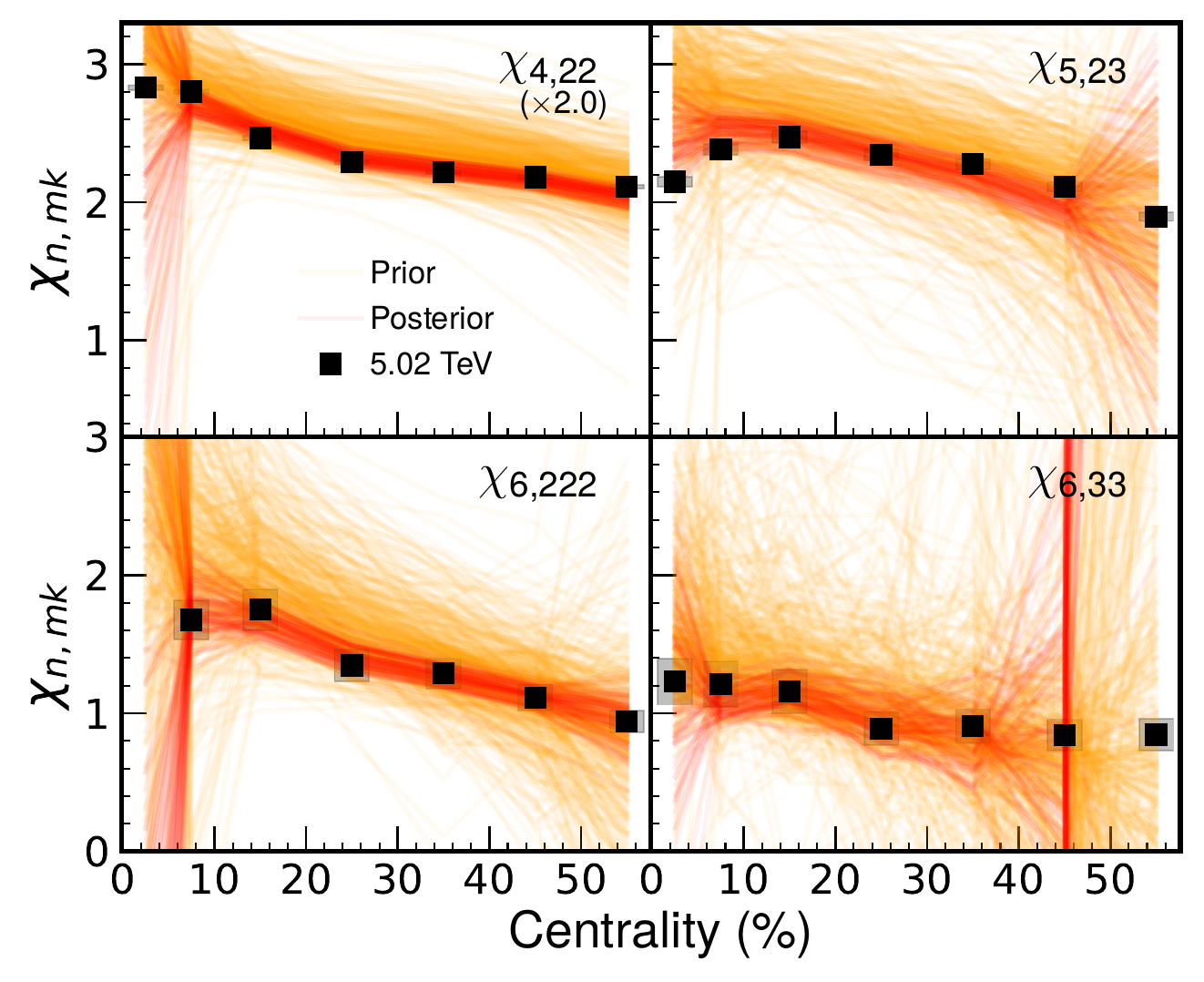}
	\caption{Design parametrizations for non linear flow mode coefficients $\chi_{n,mk}$ (in yellow) and a number of posterior sample curves as given by the emulator (in red).}
	\label{fig:chi_params}
\end{figure}

\begin{figure}[tbh!]
	\centering
		\includegraphics[width=0.82\linewidth]{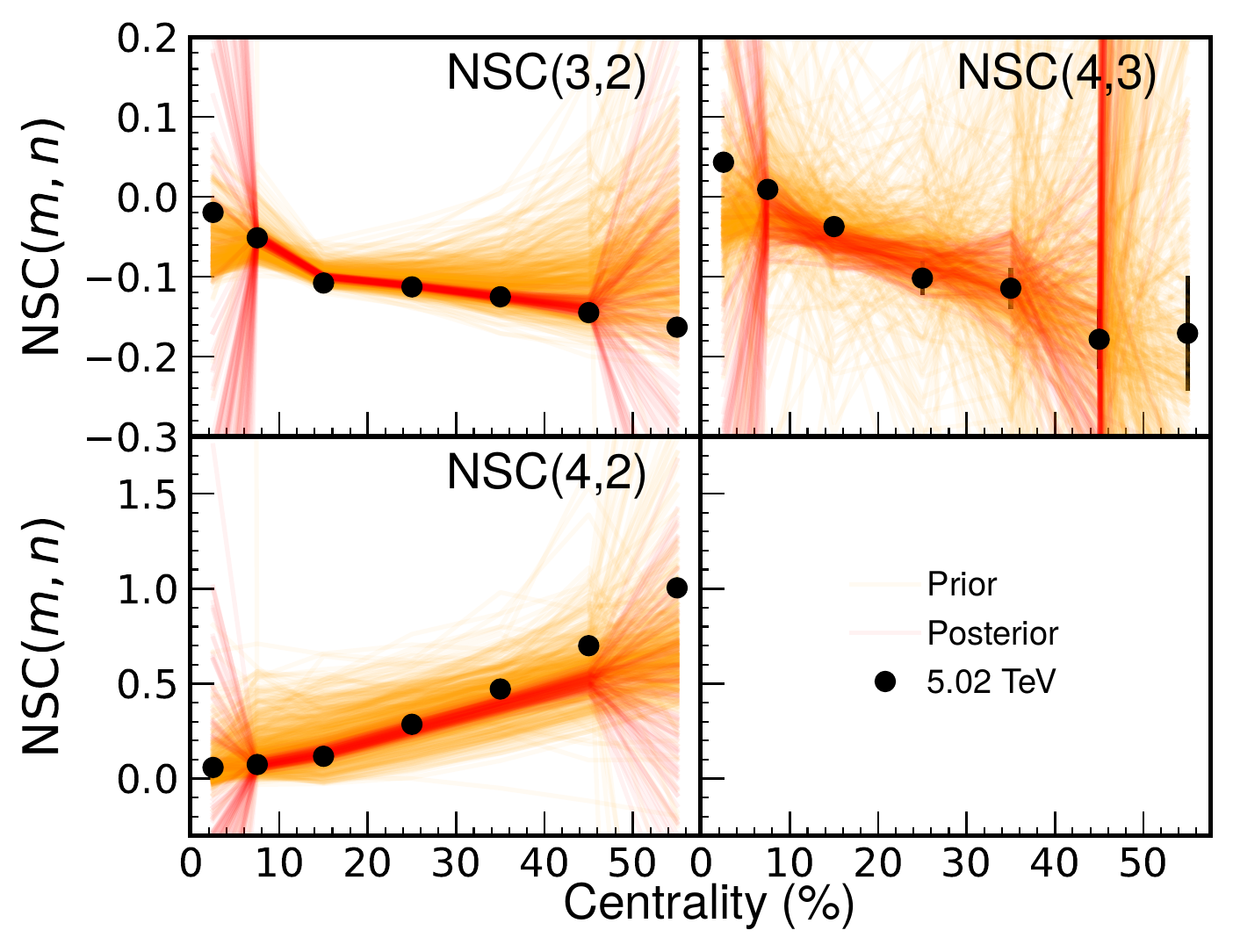}
	\caption{Design parametrizations for normalized symmetric cumulants (in yellow) and a number of posterior sample curves as given by the emulator (in red).}
	\label{fig:nsc_params}
\end{figure}

The model is calibrated to the latest Pb--Pb collision data at $\sqrt{s_\mathrm{NN}}=5.02\,\mathrm{TeV}$ from the ALICE Collaboration~\cite{Adam:2016ddh, Acharya:2019yoi, Acharya:2020taj, ALICE:2021adw}.
Figures~\ref{fig:nch_meanpt_params}--\ref{fig:nsc_params} show the calculations of each observable using the design parametrizations obtained from the prior distribution. The yellow curves represent the calculations corresponding to each design point parametrization, which are used in training the GP emulator, whereas the red curves represent emulator predictions corresponding to random points sampled from the posterior distribution.

\section{\label{sec:results}Results}

\begin{figure}[tbh!]
	\centering
	\includegraphics[width=0.8\linewidth]{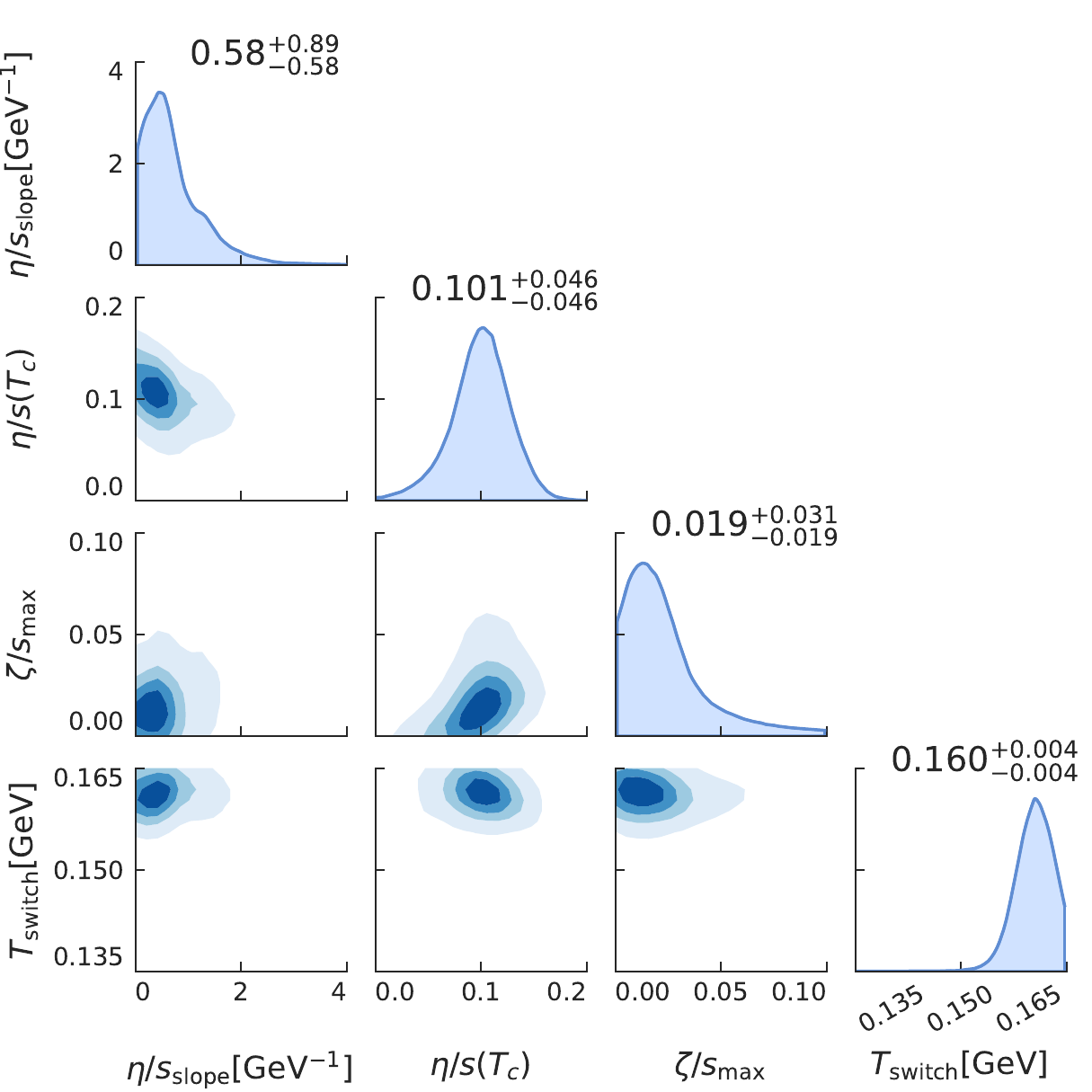} 
	\caption{Dimensionally reduced posterior probability for select transport parameters. The diagonal histograms represent the marginal distributions for the corresponding parameters. The accompanying numbers are the median values, as well as the limits of the 90\%-credibility range.}
	\label{fig:posterior_shear_bulk}
\end{figure}

Figure~\ref{fig:posterior_shear_bulk} highlights the posterior and marginal distributions for select components of the transport parameters. The primary components, $\eta/s$ slope, $\eta/s(T_c)$, $(\zeta/s)_{\max}$ in the transport parametrizations are well constrained. The initial condition parameters are well constrained within the narrow prior range.

\begin{figure}[tbh!]
	\centering
	\includegraphics[width=0.8\linewidth]{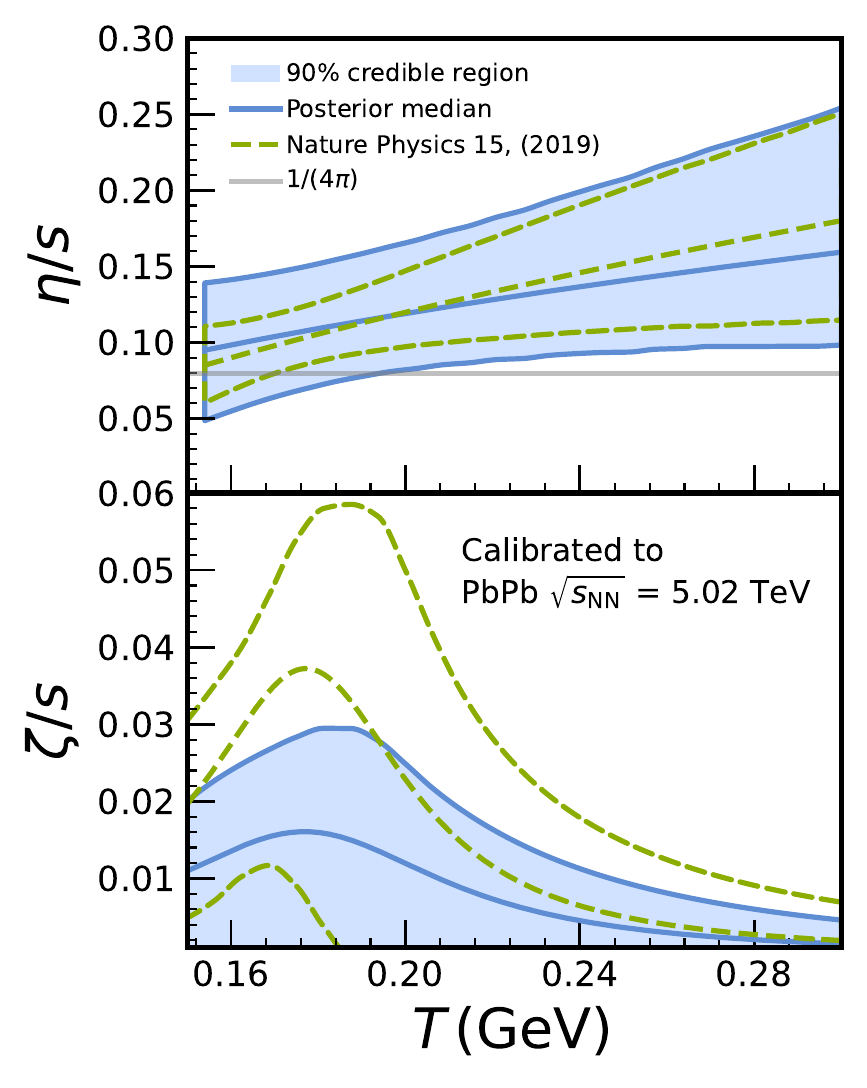}
	\caption{The 90\%-credibility region for the shear (top) and bulk (bottom) viscosity to entropy ratio is given as a blue band. The blue line represents the median of the credible range. The MAP parametrization from~\cite{Bernhard2019} as well as the corresponding 90\%-credibility range are plotted as green dashed curves.}
	\label{fig:regions}
\end{figure}

Figure~\ref{fig:regions} presents the estimated temperature dependence of $\eta/s(T)$ and $\zeta/s(T)$ according to the parametrizations from Eqs.~(\ref{eq:etaparam}) and~(\ref{eq:zetaparam}), respectively. The shaded region around the curves represents the 90\%-credibility region. 
This region reflects all uncertainties coming from the finite width of the posterior
distribution, experimental statistical and systematic uncertainties, statistical uncertainties
in model calculations, predictive uncertainty from the GP emulator, and systematic
model bias. With high probability, the true curve is located within this region.

Table~\ref{tab:optimal} presents the best-fit MAP parameters from our analysis. 
We list here the important findings:
\begin{enumerate}[(i)]
	\item While the temperature dependence of $\eta/s(T)$ is similar to what was obtained in~\cite{Bernhard2019}, the curvature of $\eta/s(T)$ is slightly stronger, resulting in lower values at higher temperatures above $T_{c}$.
	\item A notable change is the lower $(\zeta/s(T))_{\max}$ in order to reproduce the additional observables. The obtained $\zeta/s(T)$ is smaller than those found in the previous Bayesian analyses~\cite{Bernhard2019,JETSCAPE:2020mzn} where the additional observables were not included. A similar value was reported in Ref.~\cite{Nijs:2020ors}. On average this represents a value an order of magnitude lower compared to the lattice QCD calculation~\cite{Nakamura:2004sy} and the parametrizations used in~\cite{McDonald:2016vlt,Zhao:2017yhj}, where the parametrizations were tuned to simultaneously reproduce lower harmonic $v_n$ as well as the charged particle multiplicity and the low-$p_\mathrm{T}$ region of the charged hadron spectra.
	\item The switching temperature on the other hand is higher than the one found in the aforementioned studies, where on average $T_\mathrm{switch}$ is located around $\approx 0.150\,\mathrm{GeV}$. As discussed in \cite{Acharya:2017gsw,Acharya:2017zfg,Acharya:2020taj}, the additional observables, the non linear response modes and the correlations between flow harmonics are sensitive to viscous corrections to the equilibrium distribution at hadronic freeze-out~\cite{Luzum:2010ad,Luzum:2010ae,Teaney:2012ke,Yan:2015jma} and seem to prefer the higher switching temperature.
\end{enumerate}

\begin{table}
  \caption{
    \label{tab:optimal}
    The best-fit MAP parameters.
  }
  %\begin{ruledtabular}
  \begin{tabular}{|lc|lc|}
    \hline
    \multicolumn{2}{|c|}{Initial conditions} & \multicolumn{2}{c|}{Transport}\\
    Parameter         & MAP value & Parameter & MAP value \\
    \hline
    Norm              & 21.06 & $\eta/s(T_c)$       & 0.104       \\
    $p$               & 0.0077 & $(\eta/s)_\mathrm{slope}$    & 0.425 \\
    $\sigma_k$               & 0.881 & $(\eta/s)_\mathrm{crv}$ & -0.738\\
    $d_{\min}^3$               & 0.975 & $(\zeta/s)_\mathrm{peak}$    & 0.170\\
    $\tau_\mathrm{fs}$ & 0.901 & $(\zeta/s)_{\max}$ & 0.010\\
    $T_c$      & 0.147 & $(\zeta/s)_\mathrm{width}$ & 0.057\\
    $T_\mathrm{switch}$ & 0.160 & &\\
    \hline
  \end{tabular}
  %\end{ruledtabular}
\end{table}

\begin{figure}[tbh!]
	\centering
	\includegraphics[width=0.9\linewidth]{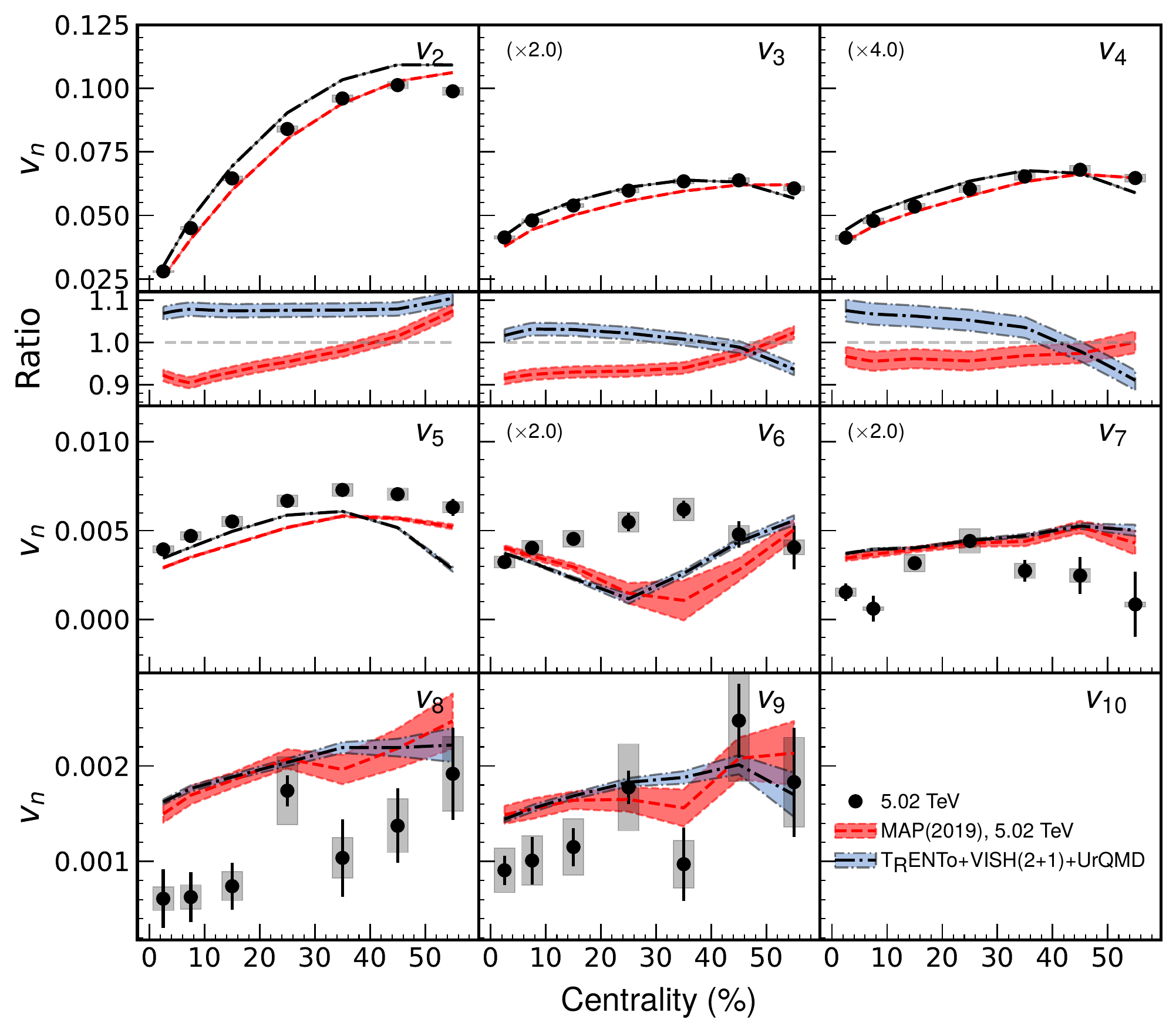}
	\caption{Flow coefficients from two hydrodynamical calculations are compared to the experimental data~\cite{Acharya:2020taj} at center-of-mass energy 5.02~TeV. The blue band is calculated with the MAP parametrization from this work, whereas the red band uses the parametrization from~\cite{Bernhard2019}.}
	\label{fig:vn_result}
\end{figure}

\begin{figure}[tbh!]
	\centering
	\includegraphics[width=0.9\linewidth]{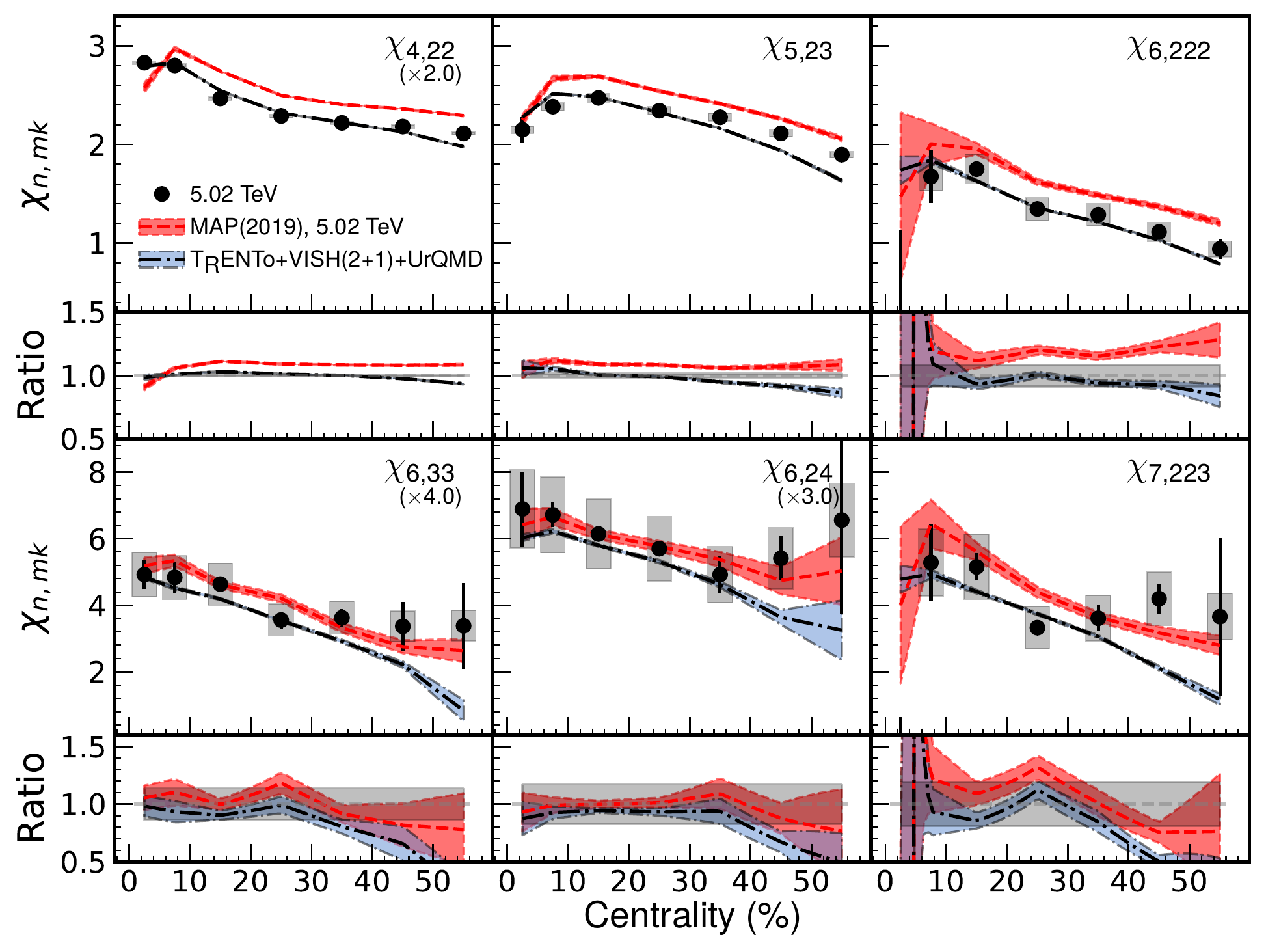}
	\caption{Non linear flow mode coefficients from two hydrodynamical calculations are compared to the experimental data~\cite{Acharya:2020taj} at center-of-mass energy 5.02~TeV. The blue band is calculated with the MAP parametrization from this work, whereas the red band uses the parametrization from~\cite{Bernhard2019}.}
	\label{fig:chi_result}
\end{figure}

\begin{figure}[tbh!]
	\centering
	\includegraphics[width=0.9\linewidth]{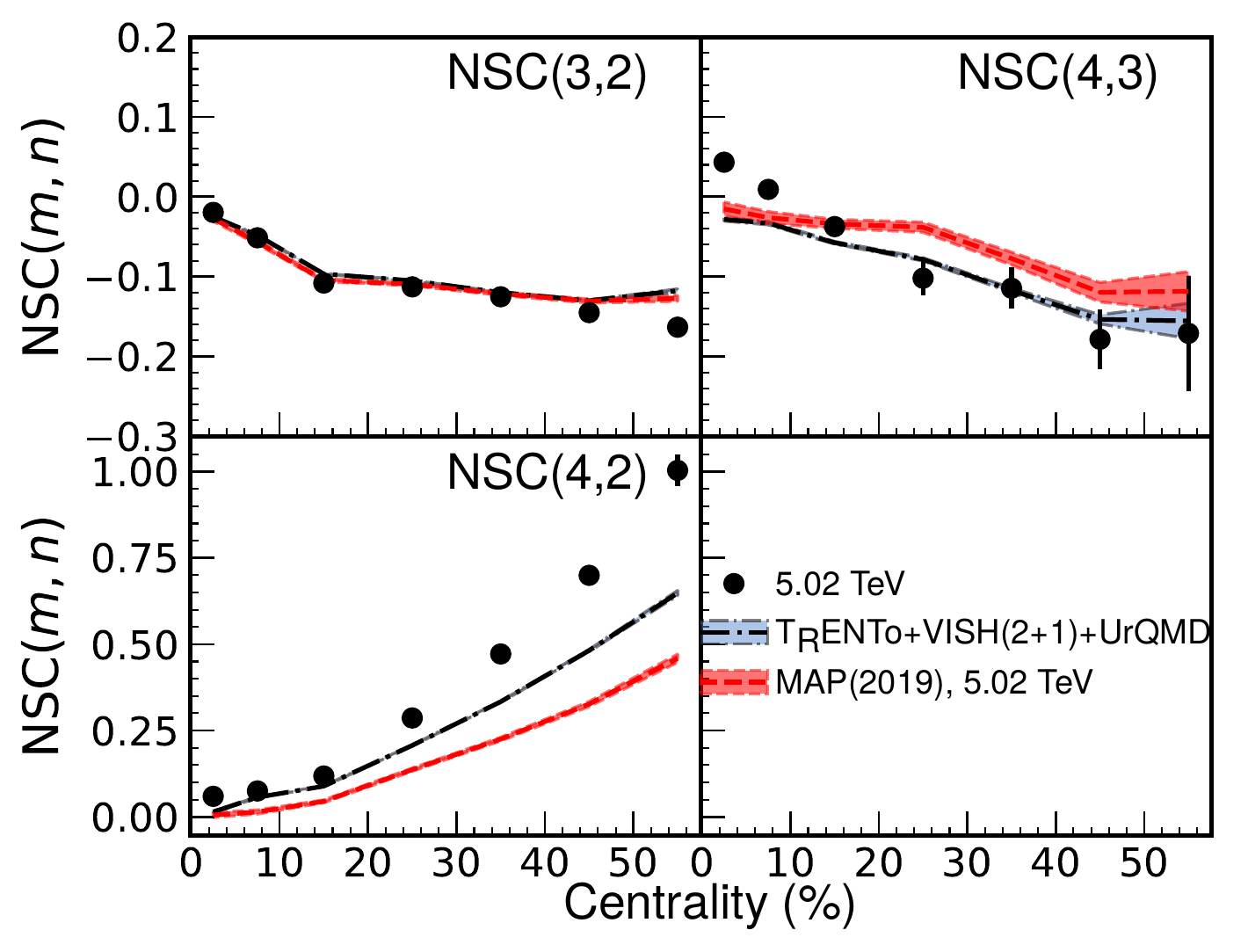}
	\caption{Normalized symmetric cumulants [$\mathrm{NSC}($m$,$n$)$) from two hydrodynamical calculations are compared to the experimental data~\cite{ALICE:2021adw} at center of mass energy of 5.02~TeV. The blue band is calculated with the MAP-parametrization from this work, whereas the red band uses the parametrization from~\cite{Bernhard2019}.}
	\label{fig:nsc_result}
\end{figure}

We performed high-statistics hydrodynamic calculations with the new parametrization. Figures~\ref{fig:vn_result} and~\ref{fig:chi_result} present the calculations for the flow coefficients $v_n$ and non linear flow mode coefficients, respectively. The $v_n$ is reproduced within 10\% agreement for $n=2$ up to $n=4$.
For the fifth harmonic, the calculations underestimate the data. The new parametrization estimates the data better in central and peripheral collisions but deviates significantly in the peripheral region. The magnitude of the successive harmonics from $v_6$ is not quite captured by the calculations within the statistical uncertainty.
Furthermore, with our new parametrization, the predictions for the non linear flow mode coefficients have also improved compared to the parametrization from~\cite{Bernhard2019}, as indicated by the ratio plots. In this case, the lower harmonic non linear flow mode coefficients are no longer overestimated, and the magnitude and centrality dependence are correctly captured. We note that the non-linear flow mode coefficients have not been included in the model calibration in~\cite{Bernhard2019}, whereas, coefficients up to $\chi_{6,33}$ and $\chi_{6,222}$ were used in this analysis. 
Figure~\ref{fig:nsc_result} presents the calculation of the normalized SC using our obtained parametrization. The performance of the new parametrization and the one from~\cite{Bernhard2019} are comparable for NSC(3,2). For NSC(4,2) and NSC(4,3), the centrality dependence is better described by the new parametrization. However, both parametrizations are unable to reproduce the strong centrality dependence of NSC(4,2), underestimating the most data points in the most peripheral collisions.
\begin{figure}[tbh!]
	\centering
	\includegraphics[width=1.0\linewidth]{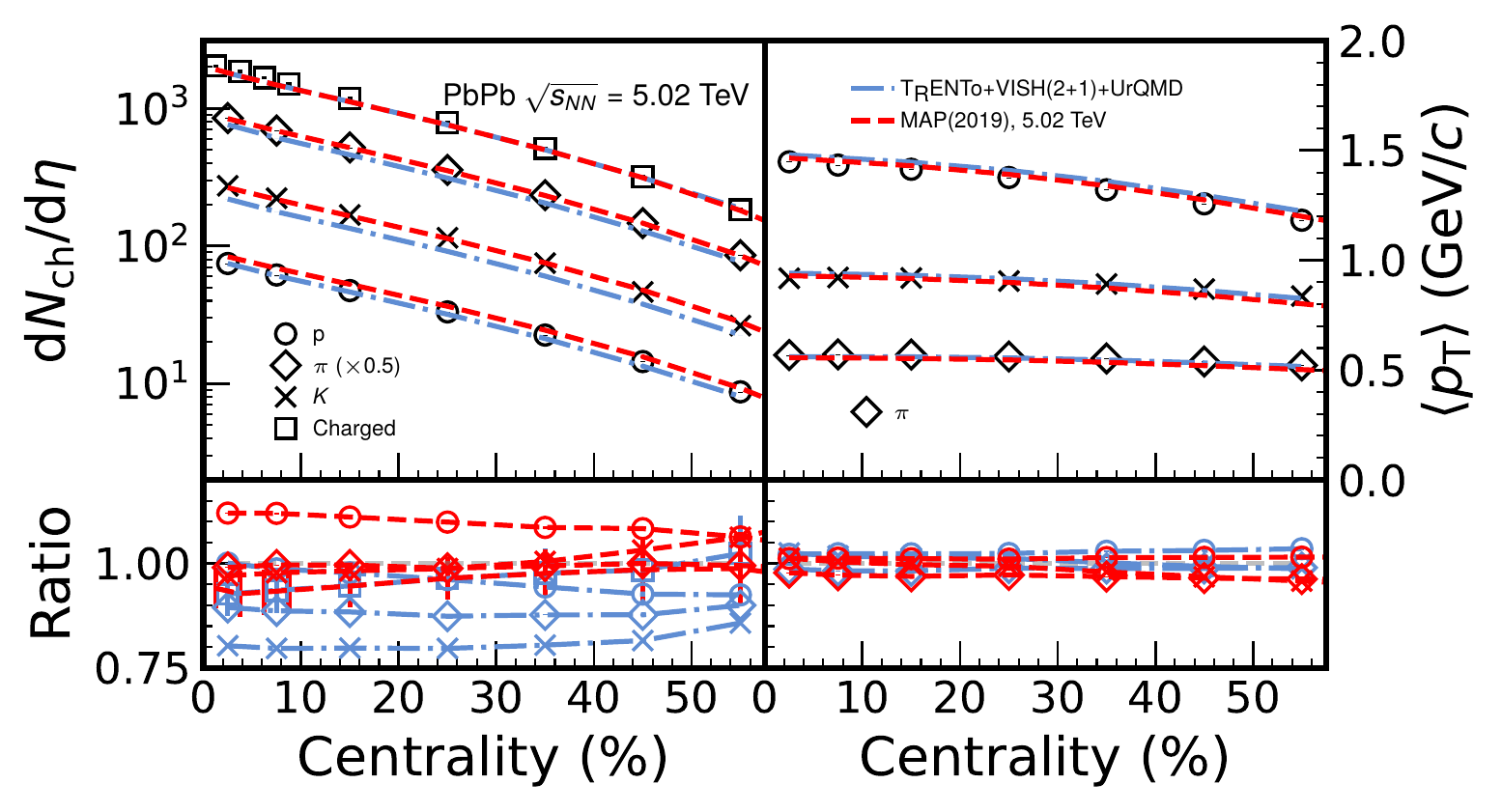}
	\caption{Charged and identified particle multiplicity and mean $p_\mathrm{T}$ from two hydrodynamical calculations are compared to the experimental data at center-of-mass energy 5.02 TeV.}
	\label{fig:mult_meanpt_MAP}
\end{figure}

\begin{figure}[tbh!]
	\centering
	\includegraphics[width=1.0\linewidth]{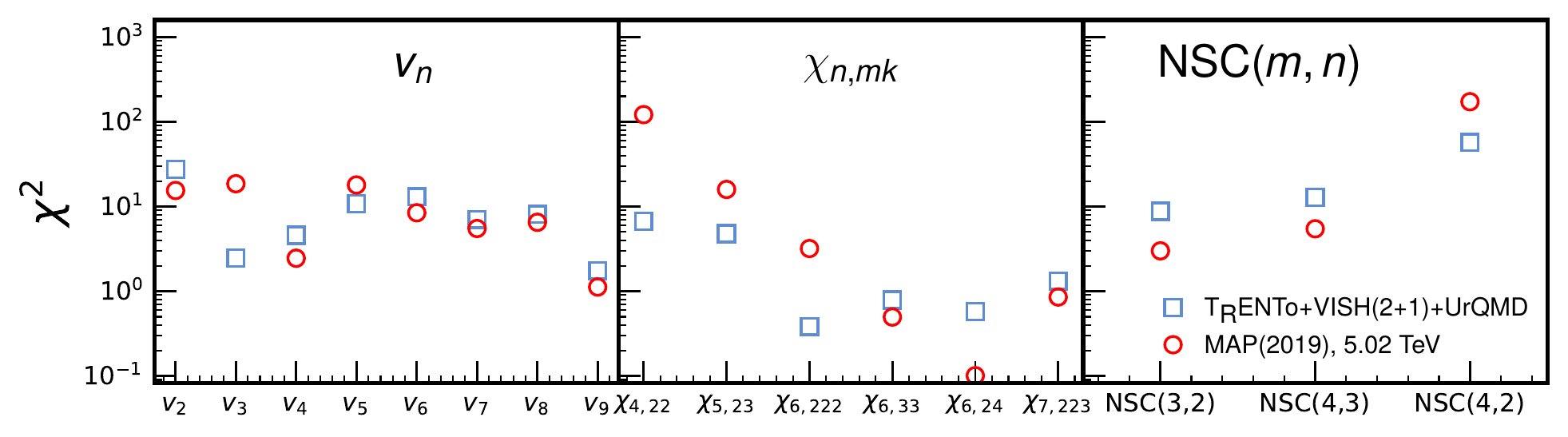}
	\caption{The $\chi^2$ values with the best-fit MAP parameters extracted from this study are compared to the ones from~\cite{Bernhard2019}.}
	\label{fig:chi2_MAP}
\end{figure}
The multiplicity and the mean $p_\mathrm{T}$ calculations are compared to the results from~\cite{Bernhard2019} in Fig.~\ref{fig:mult_meanpt_MAP}.
Our parametrization improves the estimate of the proton multiplicity and gives the same charged particle multiplicity for 5.02 TeV collisions, while the pion and kaon multiplicities are not in good agreement with the experimental data, as similarly found in \cite{Bernhard2019} for 2.76 TeV calculations. Interestingly, the parametrizations from~\cite{Bernhard2019} mainly utilizing 2.76 TeV data give better agreement with pions and kaons in 5.02 TeV collisions than our results while overestimating the proton yields approximately 10$\%$.
Agreement of the calculated mean $p_\mathrm{T}$ with the experimental data is good for all particle species as well as with the results from \cite{Bernhard2019} for both beam energies.
Refining this analysis by including low beam energy data in the future will help us to understand the beam energy dependence on various observables.

Finally, Fig.~\ref{fig:chi2_MAP} shows the $\chi^2$ values with the best-fit MAP parameters extracted from this study for each observable. They are compared to the ones from~\cite{Bernhard2019}. 
The $\chi^2$ values for our new calculation only seem to improve $v_3$ and $v_5$ for the $v_n$ observable. For the non linear flow mode coefficients the $\chi^2$ values are improved up to $\chi_{6,222}$ with our new parametrization, while the $\chi^2$ values for the higher harmonics are worse than in~\cite{Bernhard2019}. 
For NSC, the $\chi^2$ from the new calculation is worse for NSC(3,2) and NSC(4,3), but is improved for NSC(4,2). We note that the larger statistical error in the calculations using the parametrization of Ref.~\cite{Bernhard2019} lowers the corresponding $\chi^2$ values, slightly affecting the direct comparison between the two parametrizations.
The sign change of NSC(4,3) in most central collisions is not reproduced by the models while the beam energy-dependent magnitudes are better described with new parametrizations. We leave those differences for future research work where the present results should be refined by including experimental data from the lower energy beam data.

\begin{figure}[tbh!]
	\centering
	\includegraphics[width=1.0\linewidth]{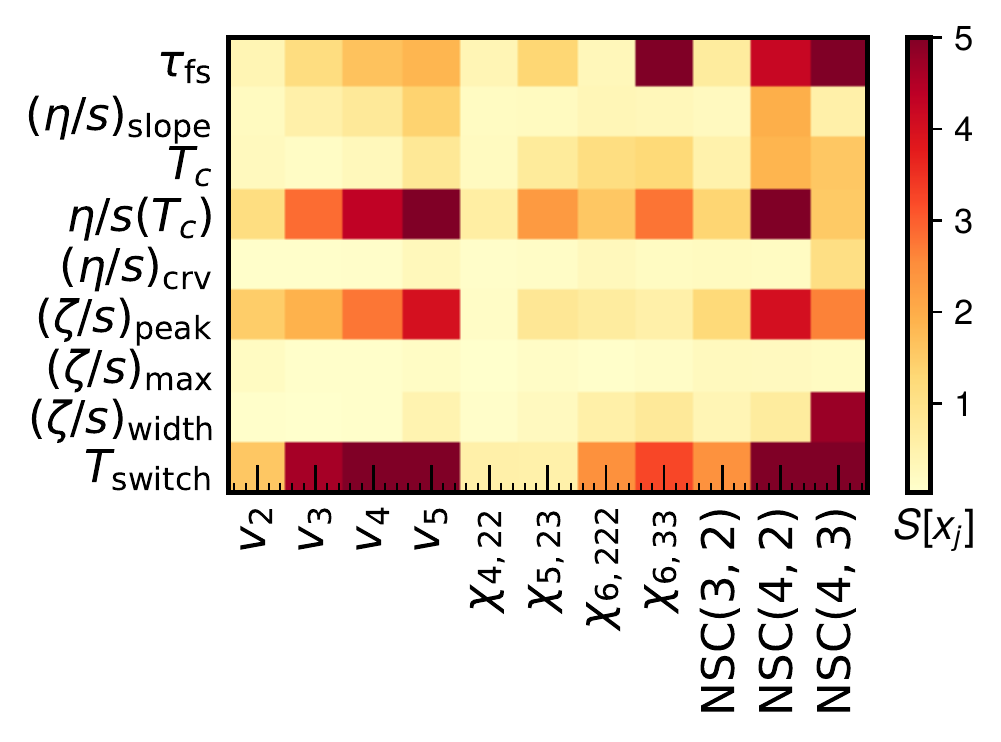}
	\caption{Sensitivity of the observables to the model parameters visualized as a colormap. The sensitivity index is averaged over four centrality classes, from 5\% to 40\%. Light yellow shades represent a very limited sensitivity or no sensitivity, whereas orange and darker red colors represent moderate or strong sensitivities to the corresponding model parameter, respectively.}
	\label{fig:sensitivity_obs}
\end{figure}
As a final study in this analysis, we conduct a simple sensitivity analysis of the included observables to the model transport parameters. The sensitivity of each observable is evaluated using the GP emulator by observing the relative difference in the magnitude of the observable between two parameter points in the parameter space. The difference can be formulated as
\begin{equation}
\Delta=\frac{|\hat{O}(\x')-\hat{O}(\x)|}{\hat{O}(\x)},
\end{equation}
where $\hat{O}(\x)$ and $\hat{O}(\x')$ represent the values of an observable at parameter points $\x$ and $\x'$, respectively~\cite{JETSCAPE:2020mzn}.

In this study, we choose a reference parameter point $\x$ to be the one representing the MAP values obtained in this analysis (see Table~\ref{tab:optimal}). To probe the sensitivity of a parameter $j$, another point is defined as $\x'=(x_1,x_2,\dots,(1+\delta)x_j,\dots,x_p)$, where $\delta$ is a small value representing a percentile change in the parameter space. We have used a value $\delta=0.1$, although larger values were observed to yield similar results.

We then calculate a final sensitivity index for each observable and parameter pair in various centrality classes as
\begin{equation}
    S[x_j]=\Delta/\delta.
\end{equation}
Figure~\ref{fig:sensitivity_obs} presents the evaluated sensitivity for each observable against the transport parameters. The sensitivity was evaluated over four centrality classes from 5\% to 40\% and averaged for the final plot. We did not observe large differences in the sensitivity between the individual centrality classes.

For $v_n$, we can verify a known fact that the sensitivity of the flow coefficients is generally very limited to the temperature dependence of $\eta/s(T)$~\cite{Niemi:2015qia}, although as expected, the sensitivity to the average $\langle\eta/s\rangle$, in this case, represented by $\eta/s(T_c)$, is very strong, and increasing at higher harmonics. The sensitivity of the $v_n$ to the $(\zeta/s)_\mathrm{peak}$ is visible, and also in this case the higher harmonics provide stronger constraints. Based on previous studies, the non-linear flow mode coefficients $\chi_{n,mk}$ are known to be sensitive to $\eta/s(T)$ at the freeze-out temperature. This is reflected by the observed sensitivity to $\eta/s(T_c)$ as well as $T_c$. By far, the normalized symmetric cumulants provide the strongest constraints to the temperature dependence of $\eta/s(T)$. This is confirmed by higher sensitivity for the other components of $\eta/s(T)$, and not only $\eta/s(T_c)$, which is also higher.

Two other parameters have also been included in this study: the free-streaming time scale $\tau_\mathrm{fs}$ and the switching temperature $T_\mathrm{switch}$. On average, the observables are reported to be generally weakly sensitive to $\tau_\mathrm{fs}$, apart from the symmetric cumulants and $\chi_{6,33}$. Furthermore, most of the observables, such as $v_n$, $\chi_{6,mk}$ and the $\mathrm{NSC}($m$,$n$)$, are seen to be highly sensitive to the switching temperature $T_\mathrm{switch}$. In both cases, the results reported here regarding $\tau_\mathrm{fs}$ and $T_\mathrm{switch}$ are not compatible with what has been observed in~\cite{JETSCAPE:2020mzn}.

\begin{figure}[tbh!]
	\centering
	\includegraphics[width=0.68\linewidth]{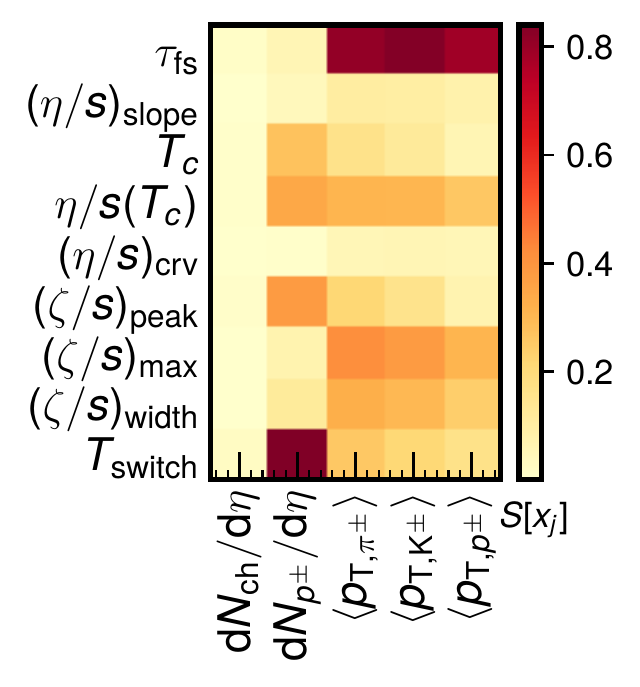} %keep it at 0.68
	\caption{Sensitivity of the mean multiplicity yields and mean transverse momenta $\langle p_\mathrm{T}\rangle$ to the model parameters.}
	\label{fig:sensitivity_energy}
\end{figure}

\begin{figure*}[tbh!]
	\centering
	\includegraphics[width=0.9\linewidth]{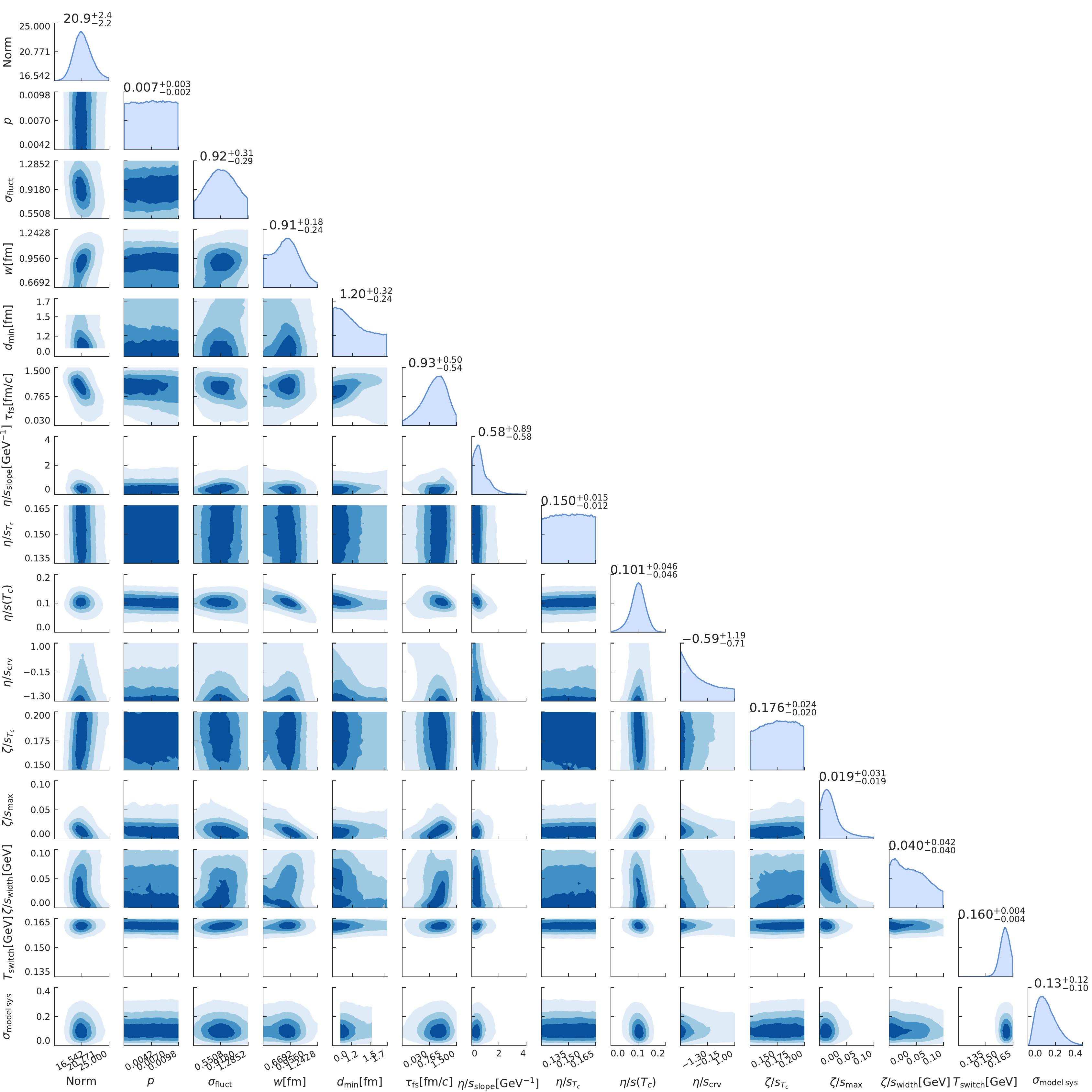}
	\caption{Dimensionally reduced posterior distribution for all parameters in the analysis. The diagonally placed histograms represent the marginal distributions for the each corresponding parameter. For each marginal distribution, a number and a range is given, denoting the median and limits of the 90\%-credibility region, respectively.}
	\label{fig:posterior}
\end{figure*}

Figure~\ref{fig:sensitivity_energy} presents the sensitivity of the multiplicity and the $\langle p_\mathrm{T}\rangle$ to the model parametrizations. Most prominently, the switching temperature affects the proton multiplicity. Furthermore, we observe a comparatively large sensitivity of $\langle p_\mathrm{T}\rangle$ to the free-streaming time scale $\tau_\mathrm{fs}$. In the case of the transport parameters, the effect on the observables is relatively small. It is observed that $\langle p_\mathrm{T}\rangle$ acts as a subtle constraint to the parameters describing the specific bulk viscosity. The posterior distribution for all parameters is shown in Fig.~\ref{fig:posterior}.

\section{\label{sec:summary}Discussions}
In summary, we performed a Bayesian analysis with the recently available data from ALICE Collaboration~\cite{Adam:2016ddh, Acharya:2019yoi, Acharya:2020taj, ALICE:2021adw} as an extension of the work~\cite{Bernhard2019}. We found that the temperature dependence of $\eta/s(T)$ is similar to what was obtained in~\cite{Bernhard2019} and that the curvature of $\eta/s(T)$ above $T_c$ is slightly lower at higher temperatures, showing weak temperature dependence of $\eta/s$. Notable changes include the lower $(\zeta/s(T))_{\max}$ and the higher switching temperature $T_\mathrm{switch}$ to reproduce additional observables such as symmetric cumulants and non-linear flow coefficients. However, the improved statistical uncertainties on both the experimental data and hydrodynamic calculations do not help to reduce the final credibility ranges. It is also noticeable that $v_5$ is still underestimated as observed in~\cite{Bernhard2019}. It is worthwhile to mention that the differences for $v_2$, $v_3$ and NSC(4,2) still remain about 5--10\% for 5.02~TeV. The sign change of NSC(4,3) in most central collisions is not reproduced by the models while the beam energy-dependent magnitudes are better described with new parametrizations. We leave those differences for future research work in which the present results should be refined by including the lower energy beam data. The parameter sensitivity analysis for the observables conducted in this study indicates that observables such as the symmetric cumulants and non linear flow modes provide a strong constraining power which, however, is still underutilized in~\cite{Bernhard2019} as well as the other Baysian analyses~\cite{Auvinen:2020mpc, Nijs:2020ors,JETSCAPE:2020mzn}. In our study, we confirm that the flow coefficients alongside the symmetric cumulants and non linear flow mode can provide some of the strongest constraints for the temperature dependence of $\eta/s(T)$ and $T_\mathrm{switch}$.
Improving aspects of the collision model, for example by replacing the initial state model with others like EKRT~\cite{Niemi:2015qia,Niemi:2015voa}, IP-Glasma~\cite{Schenke:2012wb}, and AMPT~\cite{Bhalerao:2015iya,Pang:2012he} with incorporation of nucleon substructure~\cite{Mantysaari:2017cni} in the initial conditions through an improved dynamical collision model before the hydrodynamic takes place~\cite{vanderSchee:2013pia,Romatschke:2017ejr}, might help to improve the understanding of the uncertainties of the extracted QGP properties and/or the model building blocks.

\acknowledgments
We would like to thank Harri Niemi, Kari J. Eskola and Sami R\"as\"anen for fruitful discussions. We thank Jonah E. Bernhard, J. Scott Moreland and Steffen A. Bass for the use of their viscous relativistic hydrodynamics softwares and their valuable comments on various processes of this work. We acknowledge Victor Gonzalez for his crosscheck for various technical parts of the event generation.
We acknowledge CSC - IT Center for Science in Espoo, Finland, for the allocation of the computational resources. 
This research was completed using $\approx 24$ million CPU hours provided by CSC. %18.7M BU / (100 (BU/h) * 100 nodes) * (100 nodes * 128 cpus) = ~24 

\nocite{*}

\bibliography{apssamp}% Produces the bibliography via BibTeX.

\end{document}